\documentclass[aps,prl,preprint,superscriptaddress,nobibnotes]{revtex4-2}
\usepackage{graphicx,xcolor}
\usepackage{amsmath}
\usepackage{bm}
\usepackage{ulem}
\usepackage{dcolumn}
\usepackage{ifthen}
\usepackage{multirow}
\usepackage{threeparttable}
\usepackage{siunitx}
\usepackage{hyperref}
\hypersetup{pdfnewwindow=true, linkcolor=blue, colorlinks=true, anchorcolor=blue, citecolor=blue, filecolor=blue,  menucolor=blue, urlcolor=blue}
\usepackage{setspace}
\usepackage{newunicodechar}

\begin{document}
\title{Emergence of a non-bulk hexagonal Fe$_2$S$_2$ single layer via phase transformation}

\author{Affan Safeer}
\affiliation{II. Physikalisches Institut, Universit\"{a}t zu K\"{o}ln, Z\"{u}lpicher Str. 77, 50937 Cologne, Germany \looseness=-1}
\author{Wejdan Beida}
\affiliation{Peter Gr\"{u}nberg Institut, Forschungszentrum J\"{u}lich, 52425 J\"{u}lich, Germany} 
\affiliation{Institute for Theoretical Physics, RWTH Aachen University, 52062 Aachen, Germany}
\author{Felix Oberbauer}
\affiliation{II. Physikalisches Institut, Universit\"{a}t zu K\"{o}ln, Z\"{u}lpicher Str. 77, 50937 Cologne, Germany \looseness=-1}
\author{Nicolae Atodiresei}
\affiliation{Peter Gr\"{u}nberg Institut, Forschungszentrum J\"{u}lich, 52425 J\"{u}lich, Germany} 
\author{Gustav Bihlmayer}
\affiliation{Peter Gr\"{u}nberg Institut, Forschungszentrum J\"{u}lich, 52425 J\"{u}lich, Germany} 
\author{Max Wolfertz}
\affiliation{II. Physikalisches Institut, Universit\"{a}t zu K\"{o}ln, Z\"{u}lpicher Str. 77, 50937 Cologne, Germany \looseness=-1}
\author{Chiara Schlichte}
\affiliation{II. Physikalisches Institut, Universit\"{a}t zu K\"{o}ln, Z\"{u}lpicher Str. 77, 50937 Cologne, Germany \looseness=-1}
\author{Wouter Jolie}
\affiliation{II. Physikalisches Institut, Universit\"{a}t zu K\"{o}ln, Z\"{u}lpicher Str. 77, 50937 Cologne, Germany \looseness=-1}
\author{Stefan Blügel}
\affiliation{Peter Gr\"{u}nberg Institut, Forschungszentrum J\"{u}lich, 52425 J\"{u}lich, Germany} 
\affiliation{Institute for Theoretical Physics, RWTH Aachen University, 52062 Aachen, Germany}
\author{Jeison Fischer}
\email{jfischer@ph2.uni-koeln.de}
\affiliation{II. Physikalisches Institut, Universit\"{a}t zu K\"{o}ln, Z\"{u}lpicher Str. 77, 50937 Cologne, Germany \looseness=-1}
\author{Thomas Michely}
\affiliation{II. Physikalisches Institut, Universit\"{a}t zu K\"{o}ln, Z\"{u}lpicher Str. 77, 50937 Cologne, Germany \looseness=-1}
%\date{\today}

\begin{abstract}

Two-dimensional materials can stabilize crystal structures that are absent from their bulk counterparts, offering opportunities for materials design. Here, we report the synthesis of a previously unknown hexagonal Fe$_2$S$_2$ single layer with $\beta$-CuI structure, a buckled layer of two vertically stacked FeS honeycomb lattices, realized by thermally induced transformation of single layer mackinawite grown on graphene/Ir(111). In situ scanning tunneling microscopy and low-energy electron diffraction reveal a transition from a tetragonal to a hexagonal lattice accompanied by distinct morphological and electronic signatures. The hexagonal Fe$_2$S$_2$ forms reproducibly upon annealing and represents a new structural motif within the Fe–S material family. First-principles calculations identify the $\beta$--CuI structure as most consistent with experiment. The calculations suggest that on-site Coulomb interactions and magnetic order are relevant to understanding the stability of the new 2D Fe--S compound. The preferred nucleation of single-layer mackinawite, despite being energetically disfavored, is speculated to result from its low edge energy, analogous to the 3D case. Our results establish Fe$_2$S$_2$ as a platform for exploring structural polymorphism in two dimensions and demonstrate that reduced dimensionality can stabilize crystal structures not accessible in bulk materials.

\end{abstract}

\maketitle
\newpage

\section{Introduction}

Iron sulfides are a well-investigated class of materials of interdisciplinary interest. Their phase diagram is rich \cite{Walder05}, as is the number of different minerals with natural occurrence \cite{Rickard07}. One of them, troilite, arrives on Earth primarily by meteorites \cite{Brett67} or from the moon \cite{Evans70}. Iron sulfides play a key role in the linked biogeochemical cycles of sulfur and iron, which are a major aspect of the evolution on earth \cite{Rickard07,Fakhraee25}. Sulfate reduction by microorganisms in the sea is a key route for mineralization of organic matter in marine sediments \cite{Jorgensen82, Jorgensen19}. The resulting sulfide reacts rapidly with iron to form iron sulfide species like mackinawite, which transforms over time to greigite and eventually to pyrite. The binding and storage of sulfur in pyrite is largely responsible for an oxygen rich environment on Earth \cite{Rickard07,Fakhraee25}. The iron-sulfur world hypothesis \cite{Waechtershaeuser92,Martin03,Helmbrecht25} even proposes that early life may have formed on the surface of iron sulfide minerals. Iron sulfides have many uses and potential applications.  Among them are efficient catalysts for the hydrogen evolution reaction \cite{Wang15, Miao17, Zou18,Farhan24}, energy storage \cite{Farhan24}, and water or soil remediation \cite{Gong16, Fan17}.

One might assume that, for a relevant material system investigated thoroughly over many decades, no more surprises on the existing stable crystal structures are to be expected. However, the world of two dimensions provides a surprise in this respect. Here we report that the well-known mackinawite in single layer form (see Figure~\ref{Figure1}e) transforms by annealing into a single layer hexagonal van der Waals material of $\beta$-CuI structure representing a new structure within the Fe--S material family (see Figure~\ref{Figure2}e). The discovery of the \textit{hexagonal} phase is the more remarkable, as tetragonal bulk mackinawite is known to be a metastable Fe--S compound that transforms into thermodynamically stable greigite or pyrite, which are, however,  \textit{cubic} Fe--S crystal structures \cite{Rickard07}.

Bulk mackinawite is the major Fe--S compound precipitated in aqueous solutions \cite{Rickard07}. The preferential precipitation of the thermodynamically less stable mackinawite was speculated to be due to its lower surface energy which in turn lowers its nucleation barrier \cite{Son22}, but also kinetic reasons were considered. Nearly stoichiometric bulk mackinawite crystals are superconducting below $T_\mathrm{C} = 5$\,K \cite{Lai2015, Yang16}. The magnetism of bulk mackinawite remains ambiguous. A plausible interpretation is that strong spin fluctuations down to low temperature mask the antiferromagnetic order predicted by density functional theory (DFT), rendering it undetectable to Mößbauer spectroscopy \cite{Kwon11,Koteski17}.
Strained \cite{Zhao2017, Shigekawa19} and unstrained \cite{Lin2018} single layer mackinawite was grown on SrTiO$_4$ and graphene (Gr), respectively, but no superconductivity could be detected. 

The Fe--S bulk minerals with hexagonal or monoclinic crystal structure -- troilite, pyrrhotite, and smythite -- are variants of the NiAs structure. Troilite, the end member of Fe$_{1-x}$S pyrrothite group, exhibits the coupled structural, magnetic and electronic Morin transition around 415\,K, where the structure changes from troilite (space group $P\Bar{6}2c$) to plain NiAs structure (space group $P6_3/mmc$), the antiferromagnetic order from in-plane to out-of-plane, and the electronic structure from insulating to metallic \cite{Li96,Bansal20}. Moreover, troilite was shown to possess a spontaneous Hall effect at room temperature \cite{Takagi25}. In recent years, thin films and 2D layers of hexagonal Fe--S compounds were grown in the NiAs structure \cite{Lin2018,Zhang23,Lin2018,Zhou2023} and as the van der Waals material T-FeS$_2$ (space group $P\bar{3}m1$)  \cite{Zhou2023,Prabhu2025}.

A single layer compound in the hexagonal $\beta$-CuI structure was first realized for Mn$_2$Se$_2$ \cite{Aapro2021} and later also found for Mn$_2$Te$_2$ \cite{Cuxart2025}. As usual for single layer 2D materials \cite{Haastrup_2018,Gjerding_2021}, here and in the following, we indicate the number of each atom species per unit cell in the stoichiometric formula with subscripts. The experimental finding inspired substantial theoretical work \cite{Wang2023, Qayyum24, Long25, Li26}, as antiferromagnetic order of the two metal planes within a single layer would give rise to strong intrinsic nonlinear or (in the presence of an orthogonal electric field) anomalous Hall effects. Measuring the Hall conductivity would allow efficient and fast detection of the Néel vector, which is a key issue in potential antiferromagnetic data storage devices. 

Mostly in the context of high-throughput studies, DFT calculations have already explored some properties of the Fe--S compound in the $\beta$-CuI structure \cite{Li2022,Liang24,Liu2024,Wang26}, hereafter referred to as h-Fe$_2$S$_2$. Assuming ferromagnetic order in the single layer, the authors identify a large spin-orbit gap and a topological edge state, which can be exploited for catalysis \cite{Wang26} or, in antiferromagnetically coupled multilayers, for tuning the quantum anomalous Hall effect \cite{Liang24,Liu2024}.

The manuscript first describes the structure of single layer mackinawite and of the new h-Fe$_2$S$_2$. Subsequently, we investigate stability and the transformation of single-layer mackinawite by annealing and analyze the effect of growth temperature on phase competition. Lastly, the stability of h-Fe$_2$S$_2$ is explored by DFT calculations before our results are discussed. 

\section{Results and discussion}

\subsection{The structure of t-Fe$_2$S$_2$} 

\begin{figure}[hbt!]
\includegraphics[width=\textwidth]{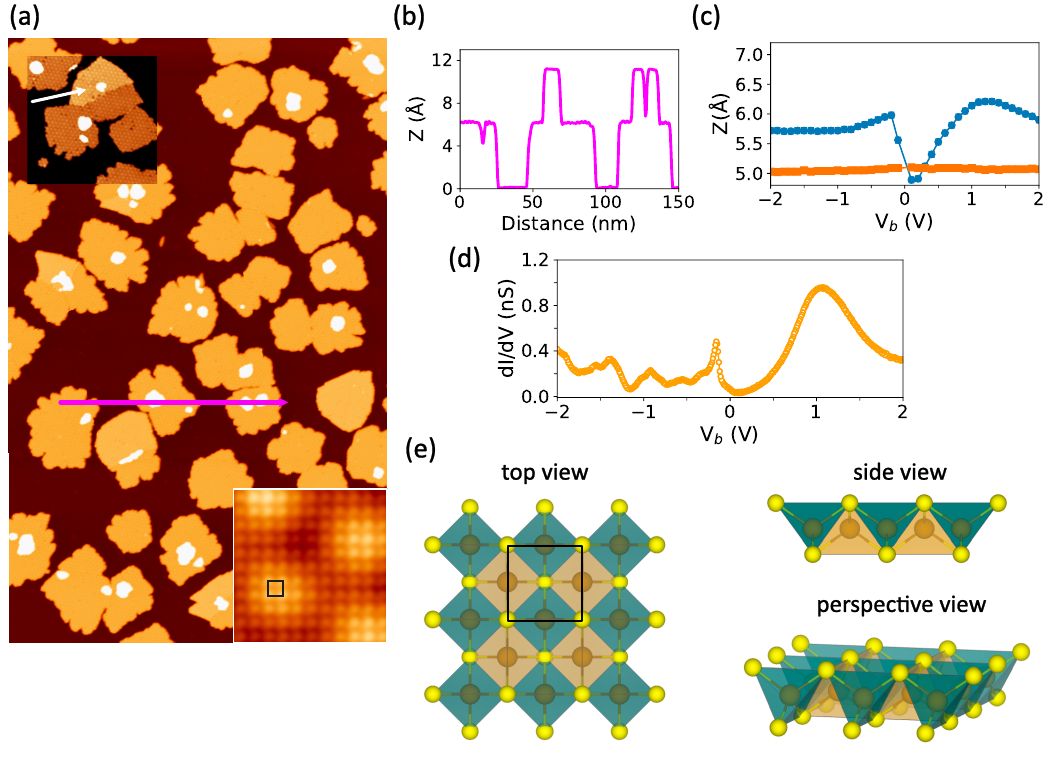}
\caption{Tetragonal single layer mackinawite (t-Fe$_2$S$_2$). (a) Overview STM topograph ($V_\mathrm{b} = 2$\,V, $I_\mathrm{t} = 50$\,pA, 400\,nm $\times$ 250\,nm) after growth at 350\,K. Inset in lower right corner is atomic resolution topograph ($V_\mathrm{b} = 100$\,mV, $I_\mathrm{t} = 1$\,nA, 4\,nm $\times$ 4\,nm) of t-Fe$_2$S$_2$. Unit cell is indicated by black square. Brightness modulation is due to Gr/Ir(111) substrate moiré. Inset in upper left corner highlights with white arrow slightly higher area of h-Fe$_2$S$_2$. (b) STM line profile along magenta line in (a). (c) Apparent height of first layer with respect to Gr (blue dots) and of second layer with respect to first layer (orange dots) of t-Fe$_2$S$_2$ as a function of sample bias $V_\mathrm{b}$ for $I_\mathrm{t} = 50$\,pA. (d) $\mathrm{d}I/\mathrm{d}V$ point spectrum ($V_{st} =2$\,V, $I_{st}=1$\,nA, $f_{mod} =677$\,Hz, and $V_{mod}= 20$\,mV) of single layer t-Fe$_2$S$_2$. (e) Ball model of t-Fe$_2$S$_2$. Unit cell with upper plane S atoms in corners is indicated by black square. Yellow balls: S; brown balls: Fe. STM and STS data taken at 1.7\,K.
}
  \label{Figure1} 
\end{figure}

An overview STM topograph of single layer mackinawite, that is, single layer tetragonal Fe$_2$S$_2$ (t-Fe$_2$S$_2$), after growth at 350\,K on Gr/Ir(111) is shown as Figure~\ref{Figure1}a. Compact islands with irregular edges and a typical diameter of $\approx 40$\,nm are observed. Several of the islands are apparently the result of coalescence as can be recognized from their larger size. Some carry a small second layer island. The total coverage of Fe$_2$S$_2$ is 48\,\% (counting the area of bilayer islands twice). The STM topograph with atomic resolution shown in the inset of Figure~\ref{Figure1}a clearly shows that the lattice is tetragonal. The lattice parameter $a_\mathrm{t}$ derived by carefully calibrated STM and LEED (see below) is $a_\mathrm{t} = 3.68\pm0.01$\,\AA, in excellent agreement with bulk mackinawite, where values $a_\mathrm{bulk} =  3.6735(4)$\,\AA~\cite{Lennie95} and $a_\mathrm{bulk} =  3.6802(5)$\,\AA~\cite{Lai2015} were reported. An STM height profile along the magenta line in Figure~\ref{Figure1}a is shown in Figure~\ref{Figure1}b. The island height is about 5.9\,\AA. A voltage dependent analysis of the apparent height of single layer islands with respect to Gr and bilayer islands with respect to single layer ones shows some bias dependence of the first, while the latter is bias-independent and can thus be identified with the geometrical layer separation (see  Figure~\ref{Figure1}c). We obtain $c_\mathrm{t} = 5.06\pm0.05$\,\AA, again in excellent agreement with $c_\mathrm{bulk} =  5.0328(7)$\,\AA~\cite{Lennie95} and $c_\mathrm{bulk} =  5.0307(7)$\,\AA~\cite{Lai2015}. 

Lastly, a large range $\mathrm{d}I/\mathrm{d}V$ point spectrum shown in Figure~\ref{Figure1}d of a single layer links the nearly vanishing differential conductance just above the Fermi level at around 100\,mV to the small apparent height at the corresponding voltage in Figure~\ref{Figure1}c. Since the tunneling probability of electrons with large parallel momentum is substantially suppressed, the vanishing conductance is still compatible with a finite density of states at this energy originating from bands located at the boundary of the first Brillouin zone. A sharp peak at about $-160$\,mV can be noticed that will be discussed below with the results of the DFT calculations.

The amount of Fe in a unit cell of t-Fe$_2$S$_2$ was analyzed by comparing the coverage of pseudomorphic Fe islands on Cu(111) with the coverage of t-Fe$_2$S$_2$ islands when identical deposition flux and time were chosen, that is, using the method as described in ref. \citenum{Safeer2025} in detail. We found a value of $2.2 \pm 0.3$ Fe atoms per unit cell. Since t-Fe$_2$S$_2$ is virtually free of defects, the number of Fe atoms per unit cell must be an integer, which is two according to our calibration. Two Fe atoms per unit cell is consistent with single layer mackinawite.

Top, side, and perspective view ball models of single layer mackinawite shown in Figure~\ref{Figure1}e highlight that each Fe$^{2+}$ ion sits in the center of a tetrahedron formed by four S$^{2-}$  ions. Each tetrahedron shares four edges with its neighbors. The metal ions are located in a single plane that is shielded by planes of S atoms. 

All experimental data (lattice symmetry, in-plane and out-of-plane lattice parameters, and the amount of Fe per unit cell) are consistent with the proposed single layer mackinawite structure. The epitaxial tetragonal Fe--S layers grown by MBE on Gr/SiC(0001) and  SrTiO$_3$(001) were similarly identified as t-Fe$_2$S$_2$ \cite{Zhao2017, Shigekawa19, Lin2018}. \\ 

\begin{figure}[t]
\includegraphics[width=\textwidth]{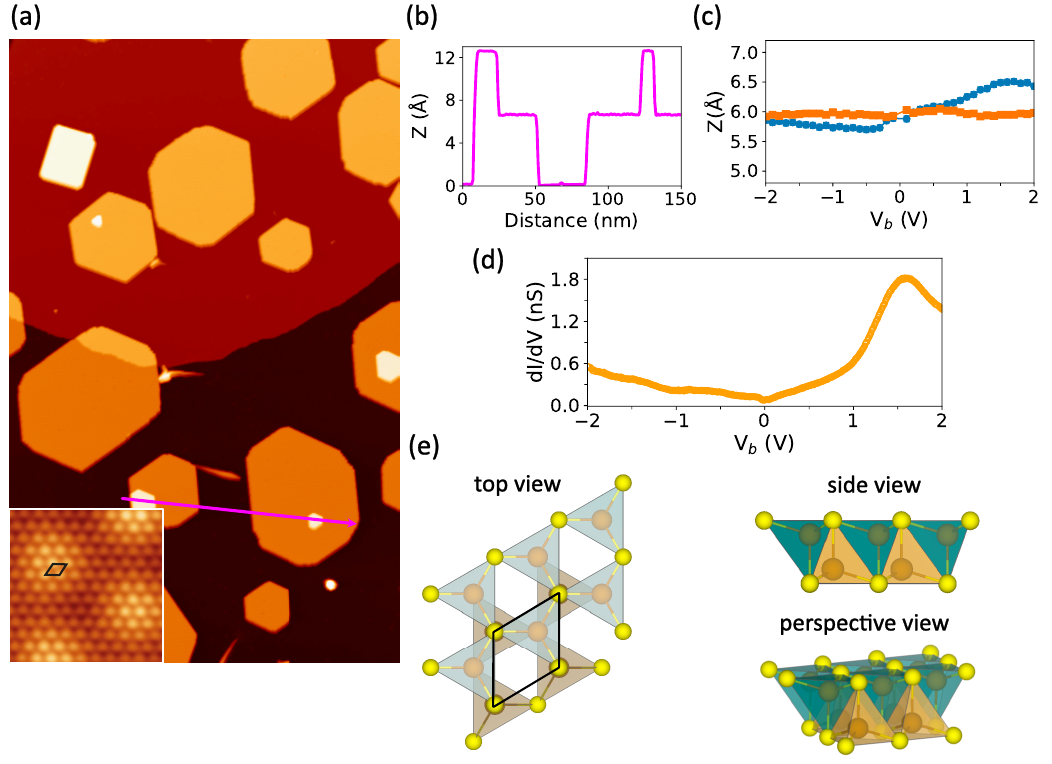}
\caption{Hexagonal single layer Fe$_2$S$_2$ in $\beta$-CuI-structure (h-Fe$_2$S$_2$). (a) Overview STM topograph after annealing sample in Figure~\ref{Figure1}a at 850\,K ($V_\mathrm{b} = 1.8$\,V, $I_\mathrm{t} = 50$\,pA, 400\,nm $\times$ 250\,nm.)  Inset in lower left corner is atomic resolution topograph ($V_\mathrm{b} = 100$\,mV, $I_\mathrm{t} = 2$\,nA, 4\,nm $\times$ 4\,nm.) of t-Fe$_2$S$_2$ ($V_\mathrm{b} = -100$\,mV, $I_\mathrm{t} = 1$\,nA, 5\,nm $\times$ 5\,nm). Unit cell is indicated by black rhombus. (b) STM line profile along blue line in (a). (c) Apparent height of first layer with respect to Gr (blue dots) and of second layer with respect to first layer (orange dots) of h-Fe$_2$S$_2$-2D as a function of sample bias $V_\mathrm{b}$ for $I_\mathrm{t} = 50$\,pA. (d) $\mathrm{d}I/\mathrm{d}V$ point spectrum ($V_{st} =2$\,V, $I_{st}=2$\,nA, $f_{mod} =677$\,Hz, and $V_{mod}= 20$\,mV) of single layer h-Fe$_2$S$_2$. (e) Top, side, and perspective view ball model of h-Fe$_2$S$_2$. Unit cell with upper plane S atoms in corners is indicated by black rhombus. Yellow balls: S; brown balls: Fe. STM and STS data taken at 1.7\,K.}
    \label{Figure2} 
\end{figure}

\subsection{The structure of h-Fe$_2$S$_2$}

After annealing the sample shown in Figure~\ref{Figure1}a to 850\,K in the absence of additional S in ultrahigh vacuum, the topography of Figure~\ref{Figure2}a is obtained. The islands now exhibit hexagonal shapes with diameters up to 70\,nm. Several islands carry small second-layer islands of threefold symmetry. The coverage has slightly decreased and is now 45\,\% averaged over several topographs. The STM topograph with atomic resolution in the inset of Figure~\ref{Figure2}a clearly shows that the lattice is hexagonal. The lattice parameter $a_\mathrm{h}$ derived by carefully calibrated STM and LEED (see below) is $a_\mathrm{h} = 3.76\pm0.02$\,\AA. An STM height profile along the magenta line in Figure~\ref{Figure2}a is shown in Figure~\ref{Figure2}b. The island height is about 6.3\,\AA. A voltage-dependent analysis of the apparent heights of single layer islands with respect to Gr and of bilayer islands with respect to single layer ones shows some bias dependence of the first, while the latter is hardly bias dependent with a scatter below $\pm 0.1$\,\AA. The expectation value of the mean is $c_\mathrm{h} = 5.97\pm0.05$\,\AA~and can be assumed to be a decent guess for the geometric layer separation (see Figure~\ref{Figure2}c). Lastly, a large range $\mathrm{d}I/\mathrm{d}V$ point spectrum of a single layer shown in Figure~\ref{Figure2}d is dominated by a strong peak in the unoccupied states at about 1.6\,V, while the DOS around the Fermi level is low. Not unexpectedly, at the bias where the $\mathrm{d}I/\mathrm{d}V$ spectrum displays a pronounced maximum, the apparent height of the first-layer islands in Figure~\ref{Figure2}c has its maximum. A $\mathrm{d}I/\mathrm{d}V$ spectrum taken on a large second-layer island is almost indistinguishable from the one shown for a single layer in Figure~\ref{Figure2}d (see Figure~S1, Supporting Information). The $\mathrm{d}I/\mathrm{d}V$ spectra for t-Fe$_2$S$_2$ in Figure~\ref{Figure1}e and for h-Fe$_2$S$_2$ in Figure~\ref{Figure2}e are quite different and leave no doubt that t-Fe$_2$S$_2$ and h-Fe$_2$S$_2$ are not only structurally but also electronically very different.

A careful inspection of the topography shown in Figure~\ref{Figure1}a after 350\,K growth reveals the presence of a small fraction of h-Fe$_2$S$_2$. H-Fe$_2$S$_2$ is at $V_\mathrm{b} = 2$\,V about 0.6\,\AA~higher than t-Fe$_2$S$_2$. In the upper left contrast-enhanced inset in Figure~\ref{Figure1}a an area of h-Fe$_2$S$_2$ is highlighted by a white arrow and can easily be identified by its larger height. Vice versa, even after annealing to 850\,K a small fraction of t-Fe$_2$S$_2$ is left, as easily identified by the square-shaped island in  Figure~\ref{Figure2}a, which is of bilayer height throughout. Careful determination of the coverages of t-Fe$_2$S$_2$ and h-Fe$_2$S$_2$ before and after annealing to 850\,K provides $2.1 \pm 0.1$ Fe atoms per unit cell of h-Fe$_2$S$_2$, assuming t-Fe$_2$S$_2$ has two Fe atoms per unit cell. 

Comparing the experimental data for h-Fe$_2$S$_2$ to all hexagonal DFT-calculated single-layer iron-sulfide polymorphs in the C2DB database \cite{Haastrup_2018,Gjerding_2021}, we find the best match for the $\beta$-CuI structure \cite{Sakuma88,Keen1995} (space group $P\bar{3}m1$) as depicted in the ball models of Figure~\ref{Figure2}e. Details of the comparison are provided in Note~2 and Table~S1 (Supporting Information). To further substantiate our assignment, ab initio calculations at the DFT level for h-Fe$_2$S$_2$ in the $\beta$-CuI structure were performed (see Note~3 and Figure~S2, Supporting Information). Our calculations agree perfectly with the results of the C2DB database, but also yield the interlayer spacing $c_\mathrm{h,DFT} = 5.87$\,\AA~in excellent agreement with the experimental value $c_\mathrm{h} = 5.97\pm0.05$\,\AA.

The ball model of Figure~\ref{Figure2}e shows that, as in t-Fe$_2$S$_2$, the Fe ions in h-Fe$_2$S$_2$ are also located at the center of edge-sharing tetrahedra, that is, both structures have the same building blocks. Taking the numbers from the C2DB database for reference, these building blocks are almost identical, with the same S-Fe bond distances up to 1\,\%, and the same S-Fe-S bond angles up to 5$^\circ$. 

\subsection{Transformation of t-Fe$_2$S$_2$ into h-Fe$_2$S$_2$ by annealing}

\begin{figure}[hbt!]
\includegraphics[width=\textwidth]{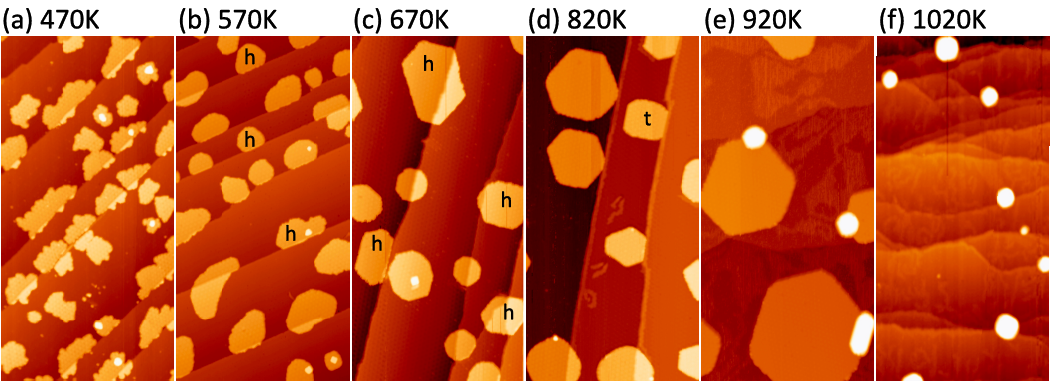}
\caption{STM annealing sequence of Fe$_2$S$_2$. After growth of Fe$_2$S$_2$ at 350\,K the sample was annealed in subsequent steps of 300\,s  at the temperatures indicated in (a) to (f). In (a), all visible islands are t-Fe$_2$S$_2$. In (b) and (c) 'h' highlights all h-Fe$_2$S$_2$ islands while in (d) 't' identifies remaining t-Fe$_2$S$_2$ islands. STM parameters: (a) $V_\mathrm{b} = 2.0$\,V,~$I_\mathrm{t} = 50$\,pA; (b) $V_\mathrm{b} = 2.0$\,V, $I_\mathrm{t} = 50$\,pA; (c) $V_\mathrm{b} = 1.3$\,V, $I_\mathrm{t} = 70$\,pA; (d) $V_\mathrm{b} = 2.5$\,mV and $I_\mathrm{t} = 35$\,pA; (e) $V_\mathrm{b} = 1.6$\,V, $I_\mathrm{t} = 60$\,pA; (f) $V_\mathrm{b} = 2.0$\,V, $I_\mathrm{t} = 70$\,pA. All topographs are 200\,nm $\times$ 100\,nm. STM data taken at 300\,K.}
  \label{Figure3} 
\end{figure}

The sequence of STM topographs in Figure~\ref{Figure3} visualizes that with a gradual increase of the annealing temperature step edge diffusion leads to more compact and regular islands shapes eventually reflecting their lattice structure. Simultaneously, with increasing temperature a transformation of t-Fe$_2$S$_2$ into h-Fe$_2$S$_2$ takes place. After the first annealing step to 470\,K, the island shapes are still irregular, and all islands in Figure~\ref{Figure3}a are still t-Fe$_2$S$_2$. Annealing to 470\,K did not induce the transformation of t-Fe$_2$S$_2$ into h-Fe$_2$S$_2$. After annealing to 570\,K a few islands transformed into h-Fe$_2$S$_2$ and are labeled 'h' in Figure~\ref{Figure3}b. First signs of a relation between structure and shape can be recognized: h-Fe$_2$S$_2$ islands develop corners fitting to  hexagonal shapes. After annealing to 670\,K in Figure~\ref{Figure3}c, the distinction between t-Fe$_2$S$_2$ and h-Fe$_2$S$_2$ is already possible just on the basis of island shapes, where the t-Fe$_2$S$_2$ islands exhibit more roundish shapes, while the 'h' islands all show hexagonal shapes. While the number of islands decreased by coarsening, the fraction of 'h' islands has increased. After annealing to 820\,K in Figure~\ref{Figure3}d the majority of islands transformed to h-Fe$_2$S$_2$. The next annealing step at 920\,K shown in Figure~\ref{Figure3}e causes the disappearance of t-Fe$_2$S$_2$. Only h-Fe$_2$S$_2$ islands are left. At the same time, each remaining island has a bright cluster of several nm height attached to it. We interpret the clusters as the result of decomposition of Fe$_2$S$_2$. The speculation on the decomposition of Fe$_2$S$_2$ is substantiated through the last annealing step to 1020\,K, after which the STM topograph of Figure~\ref{Figure3}f is recorded. Fe$_2$S$_2$ islands are absent and only clusters with a height of 3-5\,nm are found. It is plausible that the clusters are composed of bare Fe, while sulfur is either intercalated beneath Gr or sublimated to the vacuum.  

\begin{figure}[htb]
\includegraphics[width=\textwidth]{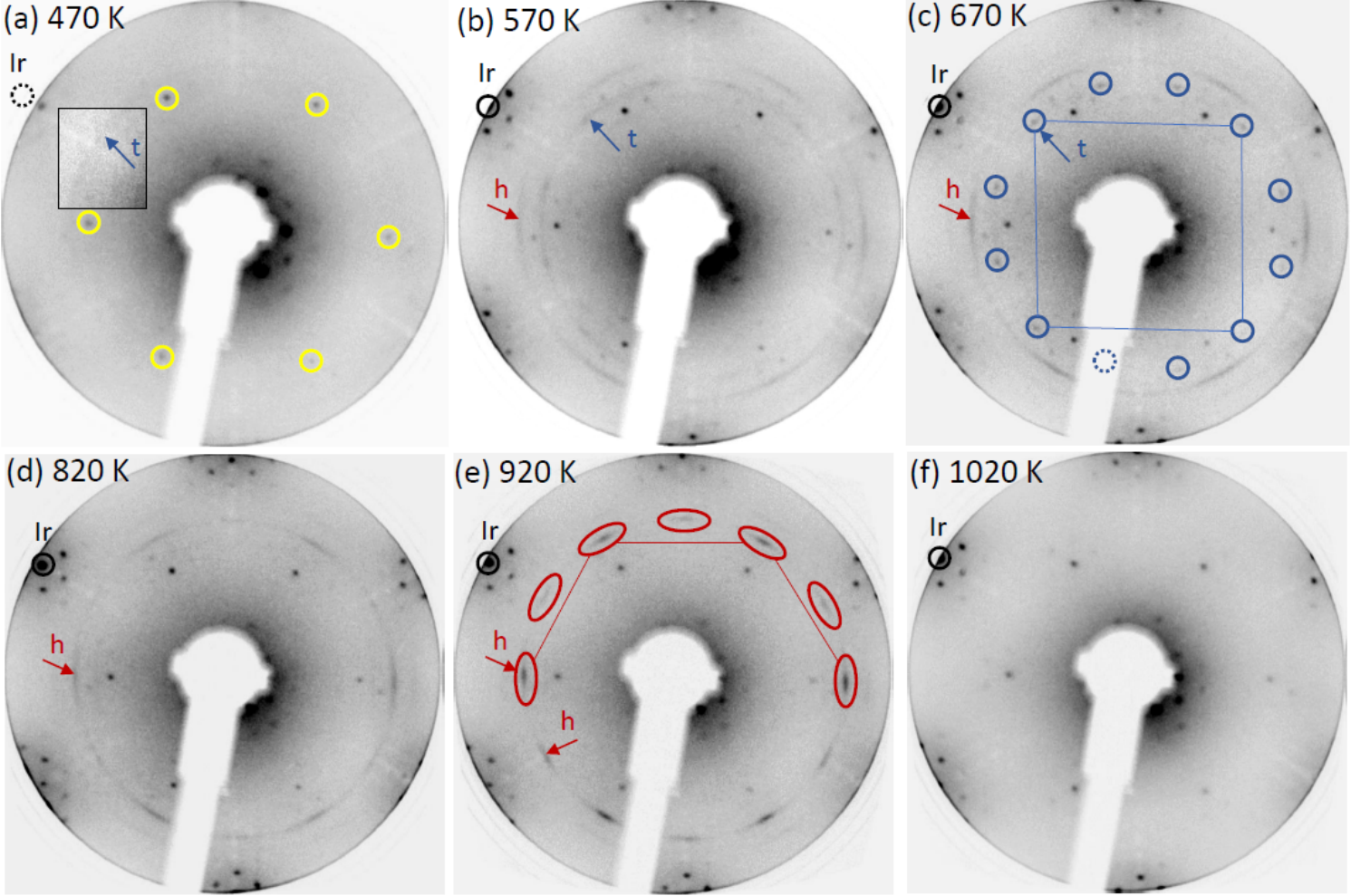}
\caption{LEED annealing sequence of Fe$_2$S$_2$. After the same annealing steps as in Figure~\ref{Figure3} LEED patterns of the sample were recorded with an electron energy of 72\,eV. The location of one first order Ir(111) spots is indicated in each LEED pattern. In (a) the first order diffraction spot of intercalated S forming a $(\sqrt3 \times \sqrt3)$-R30$^\circ$ superstructure with respect to Ir(111) are encircled yellow. The faint diffraction ring of t-Fe$_2$S$_2$ is indicated by 't' and a blue arrow in the contrast enhanced inset of (a). In (b) to (e) the diffraction rings or spots of t-Fe$_2$S$_2$ and h-Fe$_2$S$_2$ are indicated blue arrows or circles and red arrows or ellipses, respectively.}
  \label{Figure4} 
\end{figure}

A complementary view for the transformation of t-Fe$_2$S$_2$ into h-Fe$_2$S$_2$ is provided by the LEED sequence shown in Figure~\ref{Figure4}, which has the same annealing steps as the STM sequence in Figure~\ref{Figure3}. After annealing to 470\,K a faint diffraction ring (blue arrow and 't') corresponding to t-Fe$_2$S$_2$ is present in the LEED pattern of Figure~\ref{Figure4}a. Also prominent are the first order diffraction spots of S forming a $(\sqrt3 \times \sqrt3)$-R30$^\circ$ superstructure with respect to Ir(111) which are encircled yellow. These diffraction spots emerge by intercalation of S beneath Gr and their intensity depends on the perfection of the Gr layer. After annealing to 570\,K the ring for t-Fe$_2$S$_2$ becomes more pronounced and a diffraction ring for h-Fe$_2$S$_2$ appears as seen in Figure~\ref{Figure4}b. Both rings display intensity modulation with the largest intensity just in between the directions of the Ir spots. The Gr spots are aligned with the Ir spots, but at the low electron energy used outside the field of view. After annealing to 670\,K the LEED pattern in Figure~\ref{Figure4}c shows a largely unchanged segmented diffraction ring of h-Fe$_2$S$_2$, while the segmented diffraction ring of t-Fe$_2$S$_2$ in Figure~\ref{Figure4}b is now broken down into diffraction spots. The tetragonal islands take three equivalent orientations with their first order diffraction spots 15$^\circ$ off from the directions of the first order Ir and Gr spots. These orientations correspond in real space to the epiaxial relation t-Fe$_2$S$_2$-$\langle 11 \rangle \parallel$ Gr-$\langle 10 \rangle$. Annealing to 820\,K causes the tetragonal spots to largely disappear in the corresponding LEED pattern of Figure~\ref{Figure4}d. After annealing to 920\,K the h-Fe$_2$S$_2$ ring segments are shrunken down, almost to spots, and are marked by red ellipses in the upper part of Figure~\ref{Figure4}e. There are now two preferential orientations of h-Fe$_2$S$_2$. The most intense spots are 30$^\circ$ off from the directions of the first order Ir and Gr spots. In real space these h-Fe$_2$S$_2$ islands obey the epitaxial relation h-Fe$_2$S$_2$-$\langle 10 \rangle \parallel$ Gr-$\langle 11 \rangle$. The second set of six first order h-Fe$_2$S$_2$ spots is aligned to Ir and Gr, giving rise to the epitaxial relation h-Fe$_2$S$_2$-$\langle 10 \rangle \parallel$ Gr-$\langle 10 \rangle$. After annealing to 1020\,K all spots related to Fe$_2$S$_2$ have disappeared from the LEED pattern in Figure~\ref{Figure4}f, consistent with the absence of Fe$_2$S$_2$ islands in the STM picture of Figure~\ref{Figure3}f. 

LEED nicely confirms the gradual transformation of t-Fe$_2$S$_2$ into h-Fe$_2$S$_2$ and the final disappearance of h-Fe$_2$S$_2$ above 920\,K. It also highlights an aspect not obvious from the STM data, namely an improvement in the orientation alignment of t-Fe$_2$S$_2$ and h-Fe$_2$S$_2$ islands with temperature. Almost random at low temperature, at 570\,K t-Fe$_2$S$_2$ locks into one of three equivalent directions of the Gr substrate, and at 920\,K h-Fe$_2$S$_2$ aligns, though less perfectly than t-Fe$_2$S$_2$, to two different high symmetry orientations with respect to Gr. The improved alignment with increasing temperature of deposit islands with respect to the substrate is a well-known phenomenon in van der Waals epitaxy. For example, it was also found for the growth of Gr on Ir(111) by chemical vapor deposition \cite{Hattab11} or for MoS$_2$ on Gr/Ir(111) by molecular beam epitaxy \cite{Hall18}. 

\subsection{Dependence of Fe$_2$S$_2$ structure on growth temperature}
\begin{figure}[htb]
\includegraphics[width=\textwidth]{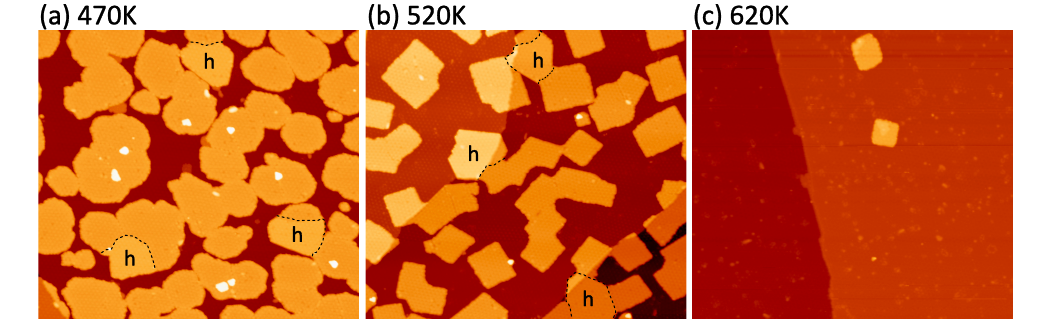}
\caption{Dependence of Fe$_2$S$_2$ structure on growth temperature. (a)-(c) Same amount of Fe deposited in identical S background pressure of $5 \times 10^{-9}$\,mbar at the temperatures indicated. The label 'h' indicates areas of h-Fe$_2$S$_2$ while all other island areas are t-Fe$_2$S$_2$. STM parameters: (a) $V_\mathrm{b} = 0.9$\,V,~$I_\mathrm{t} = 60$\,pA; (b) $V_\mathrm{b} = 1.1$\,V,~$I_\mathrm{t} = 90$\,pA; (c) $V_\mathrm{b} = 1.5$\,V,~$I_\mathrm{t} = 50$\,pA. All images 200\,nm $\times$ 180\,nm. STM data taken at 300\,K.}
  \label{Figure5} 
\end{figure}

Figure~\ref{Figure5} examines the effect of temperature on the growth process. While in Figure~\ref{Figure1}a after growth at 350\,K the islands display irregular edges, after growth at 470\,K as in Figure~\ref{Figure5}a they are more compact, and after growth at 520\,K or higher freestanding islands reflect their internal t-Fe$_2$S$_2$ structure as regular shaped squares as in Figures~\ref{Figure5}b,c. Few h-Fe$_2$S$_2$ islands are also readily identified by corners fitting to hexagonal shapes. 

A remarkable feature in the STM topograph sequence of Figure~\ref{Figure5} is the decrease in coverage with growth temperature. We rationalize this observation by considering that S re-evaporates already at room temperature from Gr/Ir(111). Increasing the growth temperature reduces the residence time $\tau$ of unreacted S according on Gr to $1/\tau = \nu_0 e^{-E_\mathrm{des}/k_\mathrm{B} T}$, where $\nu_0$ is the vibration frequency of adsorbed S towards the desorbed state, $E_\mathrm{des}$ is the activation energy for S desorption, $k_\mathrm{B}$ the Boltzmann constant, and $T$ the temperature. Consequently, the $\tau$ drops drastically with temperature. Since the flux of S is unchanged, the concentration of S on Gr at 620\,K is presumably not more sufficient to react the arriving Fe. Only a small fraction of Fe is transformed to t-Fe$_2$S$_2$, while most Fe intercalates or forms sparse Fe clusters (not present on the topograph in Figure~\ref{Figure5}c). The decrease in coverage of Fe$_2$S$_2$ at 620\,K in Figure~\ref{Figure5}c is dramatic, when referenced to growth at 420\,K in Figure~\ref{Figure5}a, but already after growth at 520\,K in Figure~\ref{Figure5}b coverage is reduced. 

A second observation in Figure~\ref{Figure5} is that the fraction of h-Fe$_2$S$_2$ does not increase significantly with increasing growth temperature. Even at 620\,K it remains below 10\,\%. It appears that Fe$_2$S$_2$ nucleates preferentially as t-Fe$_2$S$_2$, but that the energetically more favorable compound is h-Fe$_2$S$_2$, as the former irreversibly transforms into the later by annealing.

The phenomenon of preferential precipitation of metastable nanoparticulate mackinawite that transforms into more stable greigite or eventually to pyrite is well known for Fe-S compounds in the context of the global biogeochemical sulfur and iron cycles \cite{Rickard07}. The preferential precipitation of mackinawite was traced back to a lower surface energy of mackinawite compared to pyrite (FeS$_2$) \cite{Son22}. The lower surface energy of mackinawite implies a lower nucleation barrier, because the nucleation barrier is proportional to the third power of the surface energy. Nucleation rates are inversely proportional to the nucleation barrier, which implies that mackinawite nucleates preferentially due to its low surface energy, despite being energetically disfavored as a bulk material. 

The observations made here for the transformation of 2D mackinawite into the more stable h-Fe$_2$S$_2$ fit into this picture. We speculate that for this 2D case a lower step edge energy of t-Fe$_2$S$_2$ compared to h-Fe$_2$S$_2$ gives rise to enhanced nucleation of t-Fe$_2$S$_2$, but that the lower formation energy of h-Fe$_2$S$_2$ causes at sufficiently high temperature the phase transformation of metastable t-Fe$_2$S$_2$ islands into the energetically preferred h-Fe$_2$S$_2$. A prerequisite for DFT edge energy calculations that could support our speculation is an adequate description of the energetics of the 2D materials. This is difficult as discussed in the next section. 

\subsection{DFT Results: Energetics of competing Fe$_2$S$_2$ phases}

\begin{figure}[htb]
\centering
\hspace*{-0.5cm}\includegraphics[width=1.1\textwidth]{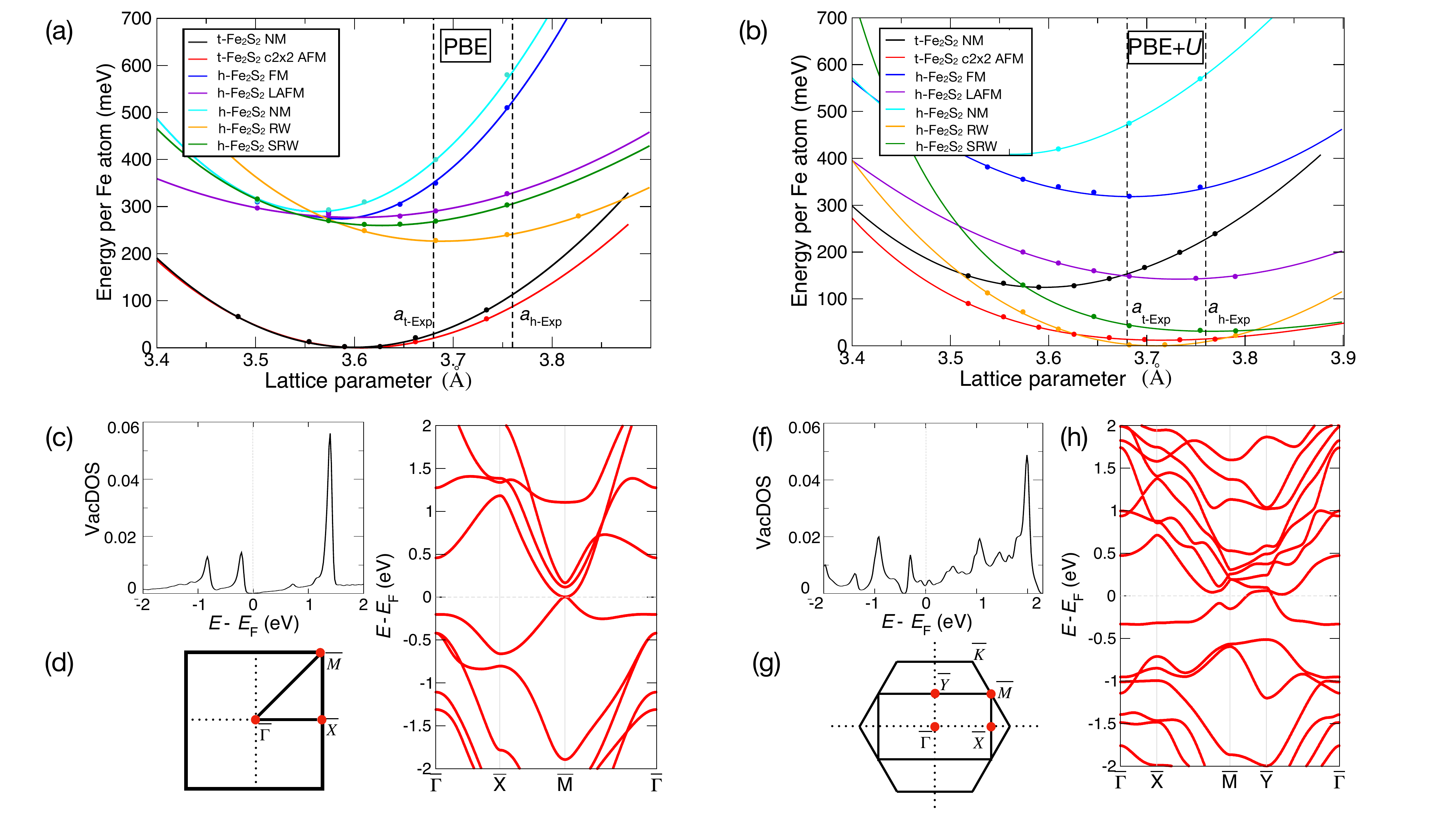}
\caption{(a), (b)  Total energy per Fe atom in (a)  with respect to $E(\text{t-Fe$_2$S$_2$ NM}, a=3.59\,\text{\AA})$ and in (b) to $E(\text{h-Fe$_2$S$_2$ RW}, a=3.72\,\text{\AA})$ as a function of in-plane lattice parameter $a$ for t-Fe$_2$S$_2$ and h-Fe$_2$S$_2$ single layers calculated within (a) the generalized gradient approximation (PBE) and (b) PBE$+U$ ($U = 1.6$~eV on Fe-$d$ states). Results are shown for non-magnetic (NM) and c(2$\times$2) checkerboard antiferromagnetic (AFM) order in t-Fe$_2$S$_2$, and NM, ferromagnetic (FM), layered AFM (LAFM), row-wise AFM (RW), and shifted row-wise AFM (SRW) order in h-Fe$_2$S$_2$. Vertical dashed lines indicate experimental lattice parameters $a_\textrm{t-Exp}=3.68$~\AA~and $a_\textrm{h-Exp}=3.76$~\AA~of the two phases. (c) Vacuum density of states (VacDOS) in states/\AA$^3$, (d) Brillouin zone, and (e) band structure of t-Fe$_2$S$_2$ in the c(2$\times$2) AFM order from PBE+$U$, along the high-symmetry lines $\overline{\Gamma}$–$\overline{\mathrm{X}}$–$\overline{\mathrm{M}}$–$\overline{\Gamma}$  of the square Brillouin zone. (f) VacDOS in states/\AA$^3$, (g) Brillouin zone, and (h) band structure of h-Fe$_2$S$_2$ in the RW order from PBE+$U$, along $\overline{\Gamma}$–$\overline{\mathrm{X}}$–$\overline{\mathrm{M}}$–$\overline{\mathrm{Y}}$–$\overline{\Gamma}$ of the rectangular Brillouin zone. The VacDOS in (c) and (f) was calculated at 4.8\,\AA\ above the S atom layer in the upper vacuum region (lower one is symmetry equivalent) and is subjected to Gaussian broadening with a half-width at half-maximum of 32\,meV.}
\label{DFT} 
\end{figure}

To explore the relative stability of the experimentally observed t- and h-Fe$_2$S$_2$ layers, we systematically investigated their total energies as a function of the lattice constant, taking into account the role of electronic correlations and magnetic order using density functional theory.

Figure~\ref{DFT}a shows the total energy per Fe atom of t-Fe$_2$S$_2$ and h-Fe$_2$S$_2$ single layers as a function of the in-plane lattice parameter, calculated using the Perdew--Burke--Ernzerhof (PBE) functional~\cite{PBE1996} for several magnetic configurations. For each in-plane lattice parameter, all atom positions were structurally optimized. Energies are referenced to nonmagnetic (NM) t-Fe$_2$S$_2$, which is the lowest-energy state in the lattice parameter range considered. The energy minimum of NM t-Fe$_2$S$_2$ is at $a = 3.59$~\AA, $-2.5$\% smaller than the experimental lattice parameter. The near degeneracy of the c($2\times 2$) AFM and NM curves reflects that, despite initializing an AFM configuration, the system converges to a nonmagnetic state for lattice constants below $\sim$3.66~\AA, with a finite magnetic moment emerging only beyond this value.

For h-Fe$_2$S$_2$, all investigated collinear magnetic configurations -- including non-magnetic (NM), ferromagnetic (FM), layered antiferromagnetic (LAFM), row-wise antiferromagnetic (RW), and shifted row-wise antiferromagnetic (SRW) order (illustrated in  Note~4 and Figure~S3, Supporting Information) -- lie energetically above the NM ground state of t-Fe$_2$S$_2$ over the full lattice-parameter range. Among these, the RW configuration of h-Fe$_2$S$_2$ represents the lowest-energy state (magnetic moment $2.50$~$\mu_\text{B}$), with a minimum located near $a = 3.70$~\AA, $ -1.6$\% smaller than the experimental lattice parameter. However, even at its minimum, RW h-Fe$_2$S$_2$ remains higher in energy than the NM t-Fe$_2$S$_2$ phase by more than $\sim 200$~meV per Fe atom.

The equilibrium lattice parameters extracted from the energy minima lie reasonably close to the experimental lattice parameters (vertical dashed lines in Figure~\ref{DFT}a). They are smaller for t-Fe$_2$S$_2$ than for h-Fe$_2$S$_2$, consistent with the lattice expansion observed experimentally upon transformation from tetragonal to hexagonal phase. 

Within PBE, t-Fe$_2$S$_2$ is predicted to be energetically favored over h-Fe$_2$S$_2$ by more than 400~meV per unit cell (200~meV per Fe atom), in contradiction to experiment, where annealing drives a transformation from t- to h-Fe$_2$S$_2$. A possible resolution is to account for the interaction of Fe$_2$S$_2$ with the Gr substrate, which is neglected in the calculations so far. If h-Fe$_2$S$_2$ binds more strongly to Gr than t-Fe$_2$S$_2$ by more than 400~meV per unit cell, the energetic preference would reverse. To test this, we performed non-magnetic DFT calculations (for details see Notes~3 and 5, Supporting Information) for t- and h-Fe$_2$S$_2$ clusters adsorbed on Gr as shown in Figure~S4 (Supporting Information). Both clusters are primarily van der Waals bound to the substrate with van der Waals gaps of 3.2\,\AA\ to 3.3\,\AA. This result is consistent with the poor orientation of both Fe$_2$S$_2$ structures, which improves only gradually with increasing annealing temperature (see Figure~\ref{Figure4}), as typical for MBE-grown van der Waals materials. 
The calculated binding energy difference is not only too small by an order of magnitude but also of the wrong sign compared to experiment. Thus, differences in binding cannot account for the discrepancy between experiment and theory. Given the predominantly van der Waals nature of the interaction, even DFT calculations including magnetic order or full Fe$_2$S$_2$ layers on Gr are not expected to qualitatively alter this conclusion. 

FeS compounds are known to exhibit strong correlations due to the interplay of electronic, lattice, and spin degrees of freedom \cite{Shimada98, Craco16, Bansal20, Li2022,Liang24,Liu2024, Wang26}. In prior studies, these correlations in the Fe 3$d$ states have been treated by an on-site Hubbard $U$ of up to 4.5~eV \cite{Shimada98, Craco16,Bansal20, Li2022, Liang24,Liu2024,Wang26} applied to the $d$ orbitals. We therefore explore the impact of DFT+$U$ on the relative energetics and subsequently also on the atomic, electronic and magnetic structure. \textit{A priori}, there is no universal value of $U$ for Fe$_2$S$_2$, as it depends also on the environment, i.e., the screening by Gr and the Ir(111) substrate. Since different structural phases may, in principle, be associated with slightly different effective $U$ values, we employ a parameter-free realization of the constrained random-phase approximation~\cite{Ersoy2011} (cRPA) and calculate the on-site Coulomb $U$ and the interorbital exchange $J$ interaction for t- and h-Fe$_2$S$_2$ single layers without substrate. The values are  $U=4.92$~eV and $J=0.63$~eV for the t-Fe$_2$S$_2$ single layer and $U=4.80$~eV and $J=0.63$~eV for  h-Fe$_2$S$_2$, both calculated for the NM states at equilibrium. We find the $J$ value is identical for both phases and the difference in $U$ is only 0.12~eV. The $U$ values are much too large as they are subject to screening due to Gr and metallic Ir(111), which could not be included because of system size. Considering a screening by the substrate to 1/3, i.e., $U\simeq 1.6$~eV, which is in the ballpark experienced for transition-metal di-oxide chains on Ir(100)~\cite{Schmitt2019,Goikoetxea2022}, the difference of $U$ between the two phases can be neglected. In order to validate our screening assumption, we follow common practice and assess measurable properties as a function of the on-site Hubbard parameter $U$ for fixed $J$ within the PBE functional approximating the exchange and correlation in DFT. We considered $U = 0$, 1.4, 1.6, and 2.8~eV, and adopt $U = 1.6$~eV throughout for consistency (see also Note 6, Supporting Information). The interorbital exchange $J$ is less ambiguous, as it is not subject to nonlocal screening; however, its value is important to correctly describe the band splitting of unoccupied minority states and the resulting bonding. Accordingly, we adopt the cRPA value $J = 0.63$ eV, which is consistent with typical values reported for Fe-based systems~\cite{Ersoy2011}. This choice of $U$ (and $J$) yields lattice constants and simulated STM spectra in good agreement with experiment, while preserving the near energetic degeneracy of the tetragonal and hexagonal phases. 

Figure~\ref{DFT}b presents the total energy per Fe atom of t- and h-Fe$_2$S$_2$ as a function of the in-plane lattice parameter for the same set of magnetic configurations as in Figure~\ref{DFT}a, but calculated within PBE+$U$ ($U=1.6$~eV, $J=0.63$~eV applied to the Fe $3d$ states).  All energies are referenced to antiferromagnetic RW h-Fe$_2$S$_2$, which is now the lowest energy structure.

For t-Fe$_2$S$_2$, the c(2$\times$2) antiferromagnetic (AFM) configuration becomes the lowest-energy state across the lattice-parameter range considered. Its minimum is located at $a =3.71$~\AA\ close to the experimental lattice parameter. The non-magnetic state lies substantially higher in energy and does not exhibit a competing minimum near the experimental lattice constant.

For h-Fe$_2$S$_2$, the inclusion of on-site Coulomb interactions significantly lowers the energies of magnetically ordered states relative to the non-magnetic configuration. Among the magnetic orders considered, the RW antiferromagnetic configuration becomes the lowest-energy state, with its energy minimum occurring near $a = 3.72$~\AA, close to the experimental lattice parameter indicated by the vertical dashed line. Relative to the RW ground state, the LAFM ($\gtrsim 136.3$~meV/Fe atom) and FM ($\gtrsim 291.1$~meV/Fe atom) states remain higher in energy. Irrespective of the choice of $U$ the previously assumed FM ground state of h-Fe$_2$S$_2$ \cite{Li2022,Liang24, Liu2024} can be ruled out even for $U=3$~eV. The NM state is strongly destabilized over the entire lattice-parameter range. 

Compared with the PBE results, the energy difference between the lowest-energy tetragonal and hexagonal magnetic configurations is essentially vanished. Near their respective equilibrium lattice parameters, the RW h-Fe$_2$S$_2$ state is energetically even slightly below the c(2$\times$2) AFM t-Fe$_2$S$_2$ state, consistent with experiment. 

The equilibrium lattice parameters of c(2$\times$2) AFM t-Fe$_2$S$_2$ and RW  h-Fe$_2$S$_2$ obtained within PBE+$U$ are shifted towards larger values relative to the PBE solutions and lie in good agreement with experiment, $+0.8$\% for t-Fe$_2$S$_2$ and $-1.1$\% for the h-Fe$_2$S$_2$.

\subsection*{DFT Results: Band structure and vacuum density of states}

To gain insight into the electronic fingerprints of the structural phases with favorable magnetic configurations, we calculated their vacuum local density of states (VacDOS) together with the corresponding band structures. Both are spin degenerate as typical for conventional N\'eel antiferromagnets subject to the combination of time reversal symmetry and sublattice translation.

For t-Fe$_2$S$_2$, Figure~\ref{DFT}c shows the calculated VacDOS. The spectrum is characterized by two pronounced peaks at $-212$~meV and $-831$~meV and zero contribution at the Fermi energy, consistent with the sharp features experimentally observed at $-160$~meV and $-920$~meV in the tunneling conductance shown in Figure~\ref{Figure1}d. The corresponding band structure, Figure~\ref{DFT}e, along the high-symmetry directions of the square Brillouin zone, Figure~\ref{DFT}d, reveals a semi-metallic behavior where valence and conduction bands touch at the Fermi-energy, consistent with the absence of tunnel conduction at about 100~meV in Figure~\ref{Figure1}d and relatively weakly dispersive spin-split Fe-$d$ bands in the vicinity of this energy window, giving rise to an enhanced density of states. Referring to one particular Fe atom, the peak at $-212$~meV arises from Fe-$d$ minority states in the vicinity of the $\overline{\Gamma}$ point and the deeper one from the Fe majority states half-way between the $\overline{\Gamma}$ and the $\overline{\mathrm{X}}$ point (for details see Figures~S5 and S6 in Note~7, Supporting Information). Concerning the peak at $-212$~meV ($-160$~meV), a closer look around $\overline{\Gamma}$ reveals that the energy dispersion $E(k_{\scriptscriptstyle{\Vert}})$ is very flat in both directions $\overline{\Gamma}-\overline{\mathrm{X}}$ and $\overline{\Gamma}-\overline{\mathrm{M}}$ and that the dispersion can be approximated by $E(k_{\scriptscriptstyle{\Vert}})=\alpha k_x^4 -\beta k_x^2 k_y^2 +\alpha k_y^4$, which explains  this peak as  a higher-order van Hove singularity with $d$-wave symmetry leading to a stronger-than-logarithmic divergence in the density of states.

In contrast, the hexagonal h-Fe$_2$S$_2$ phase exhibits a markedly different electronic structure. The calculated VacDOS (Figure~\ref{DFT}f shows multiple features distributed over a broader energy range and a reduced spectral weight near the Fermi level as well as at about $-400$~meV. The associated band structure (Figure~\ref{DFT}h), computed along the high-symmetry directions of the orthorhombic Brillouin zone  (Figure~\ref{DFT}g) corresponding to the row-wise antiferromagnetic ground state, displays a more complex dispersion with several bands crossing the Fermi level. Compared to the tetragonal phase, it is not a semi-metal and the electronic states are less localized in energy, consistent with the absence of a dominant peak close to $E_\mathrm{F}$. While the calculated VacDOS agrees with the experimental spectrum (Figure~\ref{Figure2}d) in exhibiting low spectral weight at $E_\mathrm{F}$, it does not accurately reproduce the overall spectral shape in the energy range between $E_\mathrm{F}$ and $-1$~eV. We speculate that this discrepancy arises from differences between the experimental and computed magnetic structures.

In summary, these results demonstrate that the transformation from the tetragonal to the hexagonal phase is accompanied not only by structural and magnetic changes but also by a substantial reorganization of the near-surface electronic structure that directly affects the tunneling response.

\subsection{The stability of h-Fe$_2$S$_2$}

Given that a quasi-freestanding single layer mackinawite transforms by annealing to a hitherto unknown phase of Fe$_2$S$_2$, a number of fundamental questions need to be posed. 

\textit{Why does t-Fe$_2$S$_2$ not decompose to greigite or pyrite?}
When mackinawite is reported to be metastable and to transform into the more stable phases of greigite (Fe$_3$S$_4$) and pyrite (FeS$_2$), this transformation occurs under hydrothermal conditions in the presence of dissolved sulfides \cite{Rickard07, Hunger07, Lan14, Sano20}. In our experiments the initial reaction takes place at 350\,K under sulfur rich conditions, that is, excess S supplied reevaporates to  vacuum (or intercalates under Gr). However, during annealing no additional sulfur is supplied and thus a transformation of  t-Fe$_2$S$_2$ to a more sulfur rich compound may simply be inhibited by the lack of sulfur. 
  
\textit{Why does t-Fe$_2$S$_2$ not decompose to a hexagonal NiAs-type structure?}
When Fe and S are mixed in ratio 1:1, sealed in a glass tube, and properly heated and cooled, a troilite crystal (a NiAs-type structure) results \cite{Li96, Kobayashi01}, but neither a mackinawite single crystal nor a bulk crystal in the $\beta$-CuI structure. Mackinawite is prepared by precipitation from solution as nano- or microcrystals \cite{Kwon11, Sines12, Sano20}. As we speculated, single layer mackinawite formation may be fostered by a low nucleation barrier under conditions far from equilibrium, as present in our experiments, and also for solvothermal precipitation. However, it would be expected for t-Fe$_2$S$_2$ to transform at high temperature into a NiAs-type structure, as up to now no bulk Fe--S crystals of $\beta$-CuI structure could be grown.

We can list here only a number of possible reasons, why t-Fe$_2$S$_2$ does not transform into a NiAs-type structure but into h-Fe$_2$S$_2$:

(i) In t-Fe$_2$S$_2$ Fe is tetrahedrally coordinated, as is the case in  h-Fe$_2$S$_2$.  As noted above, the tetrahedra are nearly identical in bond distances and bond angles for both structures. In NiAs-type structures the Fe ions are octahedrally coordinated. The ability to maintain the tetrahedral coordination during the solid state phase transformation may be a relevant factor, although a path of continuous transformation without breaking of the tetrahedral building blocks is not obvious.

(ii) Surface effects may be of importance. Even if energetically disfavored in bulk, a lower surface energy of h-Fe$_2$S$_2$ compared to a NiAs-type crystal structure might energetically favor the former, as long as the crystallites have a large surface to volume ratio. From this point of view, the formation of h-Fe$_2$S$_2$ instead of NiAs type Fe--S can be compared to the formation of a surface reconstruction, where atoms reorganize to a different structure to compensate for the lack of binding partners. The weak interaction between the substrate and h-Fe$_2$S$_2$ makes surface effects even more plausible. 

To be more specific, considering a transformation to NiAs platelets of similar thickness, one would end up with a layer composed of four planes, namely S-Fe-S-Fe. Surface Fe likely gives rise to a high surface energy. In fact, for 2D Cr--S compounds, the thinnest material in NiAs-type structure is Cr$_2$S$_3$-2D with S-Cr-S-Cr-S planes, that is, with metal ions fully buried under S \cite{Safeer2025}. Surface Fe could be minimized by transformation to more three dimensional crystallites but with the requirement of a more substantial reorganization. If surface effects are in fact decisive in favoring h-Fe$_2$S$_2$ in the $\beta$-CuI-structure, it might be possible to trigger transformation to NiAs-type Fe--S by annealing in sufficiently high S-pressures.

(iii) The interaction of Gr and h-Fe$_2$S$_2$ is weak and therefore interface effects are unlikely, but cannot be ruled out completely. Although binding between 
single layers of Fe$_2$S$_2$ is predominantly by van der Waals interaction, we find electron transfer from the metallic Gr substrate to Fe$_2$S$_2$ (see Note~5 and Figure~S4, Supporting Information). Specifically for single layers, as observed in our experiment, charge transfer may enter the energy balance and may favor h-Fe$_2$S$_2$ over a NiAs-type structure. However, h-Fe$_2$S$_2$ is not limited to a single layer -- also double layers maintain the $\beta$-CuI structure and display a van der Waals gap between the layers. This can be concluded from the exact agreement of the measured height difference between single and bilayer islands and the calculated interlayer spacing (see Note 3 and Figure~S2, Supporting Information). Future experiments exploring the transition from single layers to multilayers could shed light on the relative importance of surface against interface effects.

\textit{Why does the $\beta$-CuI structure appear in different transition metal -- chalcogene 2D compounds as a new phase?} The $\beta$-CuI structure has been found for h-Mn$_2$Se$_2$ single layers on NbSe$_2$ \cite{Aapro2021}, for  h-Mn$_2$Te$_2$ single layers on Gr/Ir(111), and now in the present work for h-Fe$_2$S$_2$ on Gr/Ir(111). In all three cases, the $\beta$-CuI structure is unknown for bulk crystals. The deeper reason for this coincidence is unknown. Based on our speculations above, surface and/or interface effects are a probable cause. In any case, it is to be expected that for more metal-chalcogene combinations grown by MBE the $\beta$-CuI structure will form in 2D materials.  

\textit{Comparison of experiment and DFT calculations} The experimental observation that h-Fe$_2$S$_2$ becomes the thermodynamically preferred phase upon annealing confirms that  DFT with conventional functionals such as PBE does not fully capture the relevant physical ingredients at this level of theory, as it predicts the tetragonal phase to be significantly more stable than the hexagonal phase, in contradiction to experiment. Considering that our DFT results are not part of a thermodynamic environment, additional factors may still influence the phase stability under experimental conditions. In particular, the transformation occurs at elevated temperatures and under sulfur-deficient conditions, where effects such as vibrational entropy, sulfur chemical potential, and finite-size contributions from edges may play a role. However, given the large energy differences obtained within conventional PBE, these effects alone are unlikely to account for the discrepancy without inclusion of additional electronic correlations.

To account for electronic correlations in the Fe $3d$ states, we employ a DFT+$U$ approach, which improves the description. To guide the choice of interaction parameters, we performed cRPA calculations for the isolated Fe$_2$S$_2$ single layers. These yield similar values of $U$ for both structural phases, indicating that differences in correlation strength are not the origin of their distinct stability. Taking into account the expected screening by the metallic substrate, we adopt the effective values of $U=1.6$~eV and $J=0.63$~eV. With these choices, DFT+$U$ provides a coherent description of key experimental observables. In particular, if we allow for the emergence of magnetism, the calculated lattice parameters of both t-Fe$_2$S$_2$ and h-Fe$_2$S$_2$ are in good agreement with experiment, and the relative energetics correctly reproduces the experimental finding that the hexagonal phase becomes competitive in energy. Furthermore, the calculated vacuum density of states for c($2\times 2$)-AFM t-Fe$_2$S$_2$ captures the main spectral features observed in STS, including the pronounced peak structure below the Fermi level. Finally, the theory predicts a magnetically ordered ground state for h-Fe$_2$S$_2$, consistent with the strong stabilization of magnetic configurations observed in the calculations. The exact ground state could not yet been determined as the calculated vacuum density of states for the magnetic structure of lowest energy studied does not capture the main spectral features observed in STS.

\section{Summary}

In conclusion, t-Fe$_2$S$_2$ (single layer mackinawite) and h-Fe$_2$S$_2$ (single layer Fe$_2$S$_2$ in the $\beta$-CuI structure) were synthesized on Gr/Ir(111) with molecular beam epitaxy. For t-Fe$_2$S$_2$ the lattice parameters  $(a_\mathrm{t} = 3.68\pm0.01)$\,\AA~and $c_\mathrm{t} = (5.06 \pm 0.05)$\,\AA~are in excellent agreement with bulk mackinawite. For h-Fe$_2$S$_2$ the lattice parameters are  $(a_\mathrm{h} = 3.76\pm0.01)$\,\AA~and $c_\mathrm{t} = (5.97 \pm 0.05)$\,\AA. Up to the temperature limit where Fe--S compounds could be formed, t-Fe$_2$S$_2$ nucleates with a strong preference. Upon annealing metastable t-Fe$_2$S$_2$ transforms into h-Fe$_2$S$_2$, almost phase pure when annealed at 850\,K. Above 900\,K Fe$_2$S$_2$ on Gr/Ir(111) decomposes. While initially almost random in orientation, with increasing annealing temperature t-Fe$_2$S$_2$ locks into the epitaxial relation t-Fe$_2$S$_2$-$\langle 11 \rangle \parallel$ Gr-$\langle 10 \rangle$, whereas h-Fe$_2$S$_2$ scatters in orientation around the relations h-Fe$_2$S$_2$-$\langle 10 \rangle \parallel$ Gr-$\langle 11 \rangle$ and h-Fe$_2$S$_2$-$\langle 10 \rangle \parallel$ Gr-$\langle 10 \rangle$.  Both materials are bound by van der Waals interactions to their Gr substrate. 

Density functional theory (DFT) calculations are in good agreement with experiment when a Coulomb $U$ correction is applied to the Fe $3d$ states in addition to the PBE functional. Although the chosen values ($U = 1.6$~eV and $J = 0.63$~eV) are relatively small, their inclusion improves the predicted lattice constants, the vacuum density of states of t-Fe$_2$S$_2$, and supports the assignment of h-Fe$_2$S$_2$ to the $\beta$-CuI structure. Notably, despite their modest magnitude, these corrections have a pronounced effect on the relative energetics of competing structural and magnetic phases. This highlights the crucial role of electronic correlations in achieving a comprehensive understanding of the phase behavior. 

In this context, experimental determination of the magnetic ordering in this metallic 2D system represents a key next step. With the present work, we establish a new two-dimensional van der Waals magnet, expanding the small pool of experimentally realized  systems and opening a new platform for investigating magnetism in metallic low-dimensional materials.

\section{Methods}

The experiments were carried out in two ultrahigh vacuum systems with base pressures below $1 \times 10^{-10}$\,mbar. Each system was equipped with evaporators for MBE, LEED or microchannel plate LEED (MCP-LEED), and STM operated at temperatures of 300\,K or 1.7\,K. 

Ir(111) is cleaned by cycles of noble gas sputtering and subsequent flash annealing to 1510\,K. Gr was grown by exposing the clean Ir(111) to $1\times10^{-7}$\,mbar ethylene at 300\,K for 120\,s, flash annealing to 1470\,K without ethylene, and again exposure to $3\times10^{-7}$\,mbar ethylene for 600\,s at 1370\,K \cite{vanGastel09}.

To grow Fe$_2$S$_2$, the Gr/Ir substrate is exposed at 350\,K to a Fe flux in the range of $2-5 \times 10^{16}$\,atoms m$^{-2}$ s$^{-1}$ for 300\,s from an e-beam evaporator in a S pressure of $5\times10^{-9}$\,mbar measured by a distant ion gauge. Growth does not depend critically on the S pressure. Since excess sulfur reevaporates from Gr/Ir(111), growth is under S rich conditions. Annealing is conducted in the absence of additionally supplied S. 

LEED patterns were acquired at 300\,K using electron energies ranging from 72 to 150\,eV. For MCP-LEED measurements, distortions in the reciprocal space due to the flat microchannel plate geometry were corrected using a Python script.    

STM measurements were performed at 300~K in a variable temperature and at 1.7~K in a bath cryostat STM system using tungsten tips. Constant-current topographies were recorded with sample bias $V_\mathrm{b}$ and tunneling current $I_\mathrm{t}$, as detailed in the respective figure captions. The STM images were analyzed and processed (plane subtraction, contrast correction) using WSxM software \cite{Horcas2007}. 
STS spectra were obtained at 1.7\,K with stabilization bias $V_{st}$ and stabilization current $I_{st}$ using the standard lock-in technique with modulation frequency $f_{mod}$ and modulation voltage $V_{mod}$, as specified in the captions.

Ab initio DFT simulations were performed using the all-electron full-potential linearized augmented plane-wave (FLAPW) method~\cite{Wimmer1981flapw, WeinertWimmerFreeman1982} in a two-dimensional film geometry~\cite{Krakauer1979}, with the Fe$_2$S$_2$ mono single layer embedded into two symmetry related semi-infinite vacuum regions as implemented in the \textsc{Fleur} code~\cite{fleurCode}. An augmented plane-wave basis with a cutoff parameter of \( K_{\text{max}} = 4.0\,a_\text{B}^{-1} \) and a reciprocal lattice vector cutoff of \( G_{\text{max}} = 12\,a_\text{B}^{-1} \) was employed. The muffin-tin radii were chosen as \( R_{\text{MT}}(\text{Fe}) = 2.39\,a_\text{B} \) and \( R_{\text{MT}}(\text{S}) = 1.56\,a_\text{B} \) for Fe and S atoms, respectively. Exchange--correlation effects were treated within the generalized gradient approximation (GGA) using the PBE functional~\cite{PBE1996}, supplemented by on-site Hubbard corrections with \( U = 1.6 \)~eV and \( J = 0.63 \)~eV applied to the Fe \( d \) states (GGA+\(U\)). The Hubbard $U$ and $J$ parameters for the freestanding systems (i.e., without substrate) were calculated in a methodologically consistent manner using the constrained random phase approximation (cRPA)~\cite{Ersoy2011} in combination with the FLAPW method. 2D Brillouin zone integrations were performed using Monkhorst--Pack meshes of $30 \times 30$ and $31 \times 31$ $k_{\scriptscriptstyle{\Vert}}$-points for the tetragonal and hexagonal phases, respectively. To simulate STM spectra, the local density of states in the vacuum region was evaluated using a dense $k_{\scriptscriptstyle{\Vert}}$-point mesh of $60 \times 60$. 

\section{Supporting Information}
\noindent
Supporting Information is available from the Wiley Online Library or from
the author.

\section{Acknowledgements}
\noindent
Funding from Deutsche Forschungsgemeinschaft (DFG) through CRC 1238 (project number 277146847, subprojects A01, B06 and C01) is acknowledged.

\section{Conflict of Interest}
\noindent
The authors declare no conflict of interest.

\section{Data Availability Statement}
\noindent
The data that support the findings of this study are available from the
corresponding author upon reasonable request.

\section{References}
\bibliographystyle{apsrev4-2}
\bibliography{Ref}

@article{Horcas2007,
    author = {Horcas, I. and Fernández, R. and Gómez-Rodríguez, J. M. and Colchero, J. and Gómez-Herrero, J. and Baro, A. M.},
    title = {{WSXM}: A software for scanning probe microscopy and a tool for nanotechnology},
    journal = {Review of Scientific Instruments},
    volume = {78},
    number = {1},
    pages = {013705},
    year = {2007},
    month = {01},
    issn = {0034-6748},
    doi = {10.1063/1.2432410},
    url = {https://doi.org/10.1063/1.2432410}
}

@article{Busse11,
  title = {Graphene on {{Ir}}(111): Physisorption with chemical modulation},
  author = {Busse, Carsten and Lazi{\'c}, Predrag and Djemour, Rabie and Coraux, Johann and Gerber, Timm and Atodiresei, Nicolae and Caciuc, Vasile and Brako, Radovan and N'Diaye, Alpha T. and Bl\"ugel, Stefan and Zegenhagen, J\"org and Michely, Thomas},
  journal = {Physical Review Letters},
  volume = {107},
  issue = {3},
  pages = {036101},
  numpages = {4},
  year = {2011},
  publisher = {American Physical Society},
  doi = {10.1103/PhysRevLett.107.036101},
  url = {https://link.aps.org/doi/10.1103/PhysRevLett.107.036101},
}

@article{Hall18,
doi = {10.1088/2053-1583/aaa1c5},
url = {https://doi.org/10.1088/2053-1583/aaa1c5},
year = {2018},
month = {jan},
publisher = {IOP Publishing},
volume = {5},
number = {2},
pages = {025005},
author = {Hall, Joshua and Pielić, Borna and Murray, Clifford and Jolie, Wouter and Wekking, Tobias and Busse, Carsten and Kralj, Marko and Michely, Thomas},
title = {Molecular beam epitaxy of quasi-freestanding transition metal disulphide monolayers on van der Waals substrates: a growth study},
journal = {2D Materials}
}

@Article{Safeer2025,
author={Safeer, Affan
and Ghorbani-Asl, Mahdi
and Jolie, Wouter
and Krasheninnikov, Arkady V.
and Michely, Thomas
and Fischer, Jeison},
title={Which Chromium--Sulfur Compounds Exist as 2D Material?},
journal={Advanced Functional Materials},
year={2025},
volume={35},
number={49},
pages={e00907},
issn={1616-301X},
doi={10.1002/adfm.202500907},
url={https://doi.org/10.1002/adfm.202500907}
}

@article{PR140_A1133,
  author = {W. Kohn and L. J. Sham},
  title = {Self-consistent equations including exchange and correlation effects},
  journal = {Physical Review},
  year = {1965},
  volume = {140},
  number = {4A},
  pages = {A1133--A1138},
  doi = {dx.doi.org/10.1103/PhysRev.140.A1133},
  comment = {the reference paper for KS formulation of the DFT},
}

@article{PRB89_121103R,
  author = {Ikutaro Hamada},
  title = {van der {Waals} density functional made accurate},
  journal = {Physical Review B},
  year = {2014},
  volume = {89},
  number = {12},
  pages = {121103(R)},
  doi = {dx.doi.org/10.1103/PhysRevB.89.121103},
  comment = {},
}

@article{JCP86_7184,
  author = {A. D. Becke},
  title = {On the large gradient behavior of the density functional exchange energy},
  journal = {J. Chem. Phys.},
  year = {1986},
  volume = {85},
  number = {12},
  pages = {7184--7187},
  doi = {http://dx.doi.org/10.1063/1.451353},
  comment = {},
}

@article{PRB50_17953,
  author = {Peter E. Bl\"ochl},
  title = {Projector augmented-wave method},
  journal = {Physical Review B},
  year = {1994},
  volume = {50},
  number = {24},
  pages = {17953--17979},
  doi = {10.1103/PhysRevB.50.17953},
  comment = {the reference paper for PAW method},
}

@article{PRB47_558,
  author = {G. Kresse and J. Hafner},
  title = {Ab initio molecular dynamics for liquid metals},
  journal = {Physical Review B},
  year = {1993},
  volume = {47},
  number = {1},
  pages = {558--561},
  doi = {10.1103/PhysRevB.47.558},
  url = {https://doi.org/10.1103/PhysRevB.47.558},
  comment = {reference paper for VASP},
}

@article{PRB54_11169,
  author = {G. Kresse and J. Furthm\"uller},
  title = {Efficient iterative schemes for \textit{ab initio} total-energy calculations using a plane-wave basis set},
  journal = {Physical Review B},
  year = {1996},
  volume = {54},
  number = {16},
  pages = {11169--11186},
  doi = {10.1103/PhysRevB.54.11169},
  url = {https://doi.org/10.1103/PhysRevB.54.11169},
  comment = {reference paper for VASP},
}

@article{vanGastel09,
  author = {van Gastel, R. and N'Diaye, A. T. and Wall, D. and Coraux, J. and Busse, C. and Buckanie, N. M. and Meyer zu Heringdorf, F.-J. and Horn von Hoegen, M. and Michely, T. and Poelsema, B.},
  title = {Selecting a single orientation for millimeter sized graphene sheets},
  journal = {Applied Physics Letters},
  volume = {95},
  number = {12},
  pages = {121901},
  year = {2009},
  issn = {0003-6951},
  doi = {10.1063/1.3225554},
  url = {https://doi.org/10.1063/1.3225554},
}

@Article{Lin2018,
  author = {Lin, Haicheng and Huang, Wantong and Zhao, Kun and Lian, Chaosheng and Duan, Wenhui and Chen, Xi and Ji, Shuai-Hua},
  title = {Growth of atomically thick transition metal sulfide films on graphene/6{{H-SiC}}(0001) by molecular beam epitaxy},
  journal = {Nano Research},
  year = {2018},
  day = {01},
  volume = {11},
  number = {9},
  pages = {4722-4727},
  issn = {1998-0000},
  doi = {10.1007/s12274-018-2054-4},
  url = {https://doi.org/10.1007/s12274-018-2054-4},
}

@article{Aapro2021,
  author = {Aapro, Markus and Huda, Md. Nurul and Karthikeyan, Jeyakumar and Kezilebieke, Shawulienu and Ganguli, Somesh C. and Herrero, H{\'e}ctor Gonz{\'a}lez and Huang, Xin and Liljeroth, Peter and Komsa, Hannu-Pekka},
  title = {Synthesis and properties of monolayer {{MnSe}} with unusual atomic structure and antiferromagnetic ordering},
  journal = {ACS Nano},
  volume = {15},
  number = {8},
  pages = {13794-13802},
  year = {2021},
  doi = {10.1021/acsnano.1c05532},
  url = {https://doi.org/10.1021/acsnano.1c05532},
}

@article{Cuxart2025,
  author = {G. Cuxart, Marc and Robles, Roberto and Muñiz Cano, Beatriz and Gargiani, Pierluigi and Rebanal, Clara and Di Bernardo, Iolanda and Amiri, Alireza and Calleja, Fabián and Garnica, Manuela and Valbuena, Miguel A. and L. Vázquez de Parga, Amadeo},
  title = {Emergent magnetic structures at the {{2D}} limit of the altermagnet {{MnTe}}},
  journal = {Advanced Functional Materials},
  year = {2025},
  pages = {e16924},
  doi = {https://doi.org/10.1002/adfm.202516924},
  url = {https://advanced.onlinelibrary.wiley.com/doi/abs/10.1002/adfm.202516924},
}

@Article{Li2022,
  author = {Li, Jiayu and Yao, Qiushi and Wu, Lin and Hu, Zongxiang and Gao, Boya and Wan, Xiangang and Liu, Qihang},
  title = {Designing light-element materials with large effective spin-orbit coupling},
  journal = {Nature Communications},
  year = {2022},
  day = {17},
  volume = {13},
  number = {1},
  pages = {919},
  issn = {2041-1723},
  doi = {10.1038/s41467-022-28534-y},
  url = {https://doi.org/10.1038/s41467-022-28534-y},
}

@article{Wang2023,
  title = {Intrinsic Nonlinear Hall Detection of the N\'eel Vector for Two-Dimensional Antiferromagnetic Spintronics},
  author = {Wang, Jizhang and Zeng, Hui and Duan, Wenhui and Huang, Huaqing},
  journal = {Physical Review Letters},
  volume = {131},
  issue = {5},
  pages = {056401},
  numpages = {8},
  year = {2023},
  publisher = {American Physical Society},
  doi = {10.1103/PhysRevLett.131.056401},
  url = {https://link.aps.org/doi/10.1103/PhysRevLett.131.056401}
}

@article{Lai2015,
author = {Lai, Xiaofang and Zhang, Hui and Wang, Yingqi and Wang, Xin and Zhang, Xian and Lin, Jianhua and Huang, Fuqiang},
title = {Observation of superconductivity in tetragonal FeS},
journal = {Journal of the American Chemical Society},
volume = {137},
number = {32},
pages = {10148-10151},
year = {2015},
doi = {10.1021/jacs.5b06687},
URL = {https://doi.org/10.1021/jacs.5b06687},
}

@article{Zhao2017,
doi = {10.1088/0256-307X/34/8/087401},
url = {https://doi.org/10.1088/0256-307X/34/8/087401},
year = {2017},
publisher = {Chinese Physical Society and IOP Publishing Ltd},
volume = {34},
number = {8},
pages = {087401},
author = {Zhao, Kun and Lin, Hai-Cheng and Huang, Wan-Tong and Hu, Xiao-Peng and Chen, Xi and Xue, Qi-Kun and Ji, Shuai-Hua},
title = {Molecular beam epitaxy growth of tetragonal FeS films on SrTiO3 (001) substrates},
journal = {Chinese Physics Letters},
}

@article{Haastrup_2018,
doi = {10.1088/2053-1583/aacfc1},
url = {https://doi.org/10.1088/2053-1583/aacfc1},
year = {2018},
month = {sep},
publisher = {IOP Publishing},
volume = {5},
number = {4},
pages = {042002},
author = {Haastrup, Sten and Strange, Mikkel and Pandey, Mohnish and Deilmann, Thorsten and Schmidt, Per S and Hinsche, Nicki F and Gjerding, Morten N and Torelli, Daniele and Larsen, Peter M and Riis-Jensen, Anders C and Gath, Jakob and Jacobsen, Karsten W and Jørgen Mortensen, Jens and Olsen, Thomas and Thygesen, Kristian S},
title = {The Computational 2D Materials Database: high-throughput modeling and discovery of atomically thin crystals},
journal = {2D Materials},
}

@article{Gjerding_2021,
doi = {10.1088/2053-1583/ac1059},
url = {https://doi.org/10.1088/2053-1583/ac1059},
year = {2021},
month = {jul},
publisher = {IOP Publishing},
volume = {8},
number = {4},
pages = {044002},
author = {Gjerding, Morten Niklas and Taghizadeh, Alireza and Rasmussen, Asbjørn and Ali, Sajid and Bertoldo, Fabian and Deilmann, Thorsten and Knøsgaard, Nikolaj Rørbæk and Kruse, Mads and Larsen, Ask Hjorth and Manti, Simone and Pedersen, Thomas Garm and Petralanda, Urko and Skovhus, Thorbjørn and Svendsen, Mark Kamper and Mortensen, Jens Jørgen and Olsen, Thomas and Thygesen, Kristian Sommer},
title = {Recent progress of the Computational 2D Materials Database (C2DB)},
journal = {2D Materials},
}

@article{Lennie95, 
title={Synthesis and Rietveld crystal structure refinement of mackinawite, tetragonal FeS}, 
volume={59}, 
DOI={10.1180/minmag.1995.059.397.10}, 
url={https://doi.org/10.1180/minmag.1995.059.397.10},
number={397}, 
journal={Mineralogical Magazine}, 
author={Lennie, A. R. and Redfern, S. A. T. and Schofield, P. F. and Vaughan, D. J.}, 
year={1995}, 
pages={677–683}
}

@article{Kwon11,
  title = {Magnetic ordering in tetragonal FeS: Evidence for strong itinerant spin fluctuations},
  author = {Kwon, Kideok D. and Refson, Keith and Bone, Sharon and Qiao, Ruimin and Yang, Wan-li and Liu, Zhi and Sposito, Garrison},
  journal = {Physical Review B},
  volume = {83},
  issue = {6},
  pages = {064402},
  numpages = {7},
  year = {2011},
  month = {Feb},
  publisher = {American Physical Society},
  doi = {10.1103/PhysRevB.83.064402},
  url = {https://link.aps.org/doi/10.1103/PhysRevB.83.064402}
}

@article{Sakuma88,
author = {Sakuma ,Takashi},
title = {Crystal Structure of β-CuI},
journal = {Journal of the Physical Society of Japan},
volume = {57},
number = {2},
pages = {565-569},
year = {1988},
doi = {10.1143/JPSJ.57.565},
URL = {https://doi.org/10.1143/JPSJ.57.565}
}

@article{Zhang23,
title = {Controllable synthesis of two-dimensional magnetic FeS nanoplates with high Curie temperature and unique magnetotransport properties},
journal = {Nano Today},
volume = {49},
pages = {101794},
year = {2023},
issn = {1748-0132},
doi = {https://doi.org/10.1016/j.nantod.2023.101794},
url = {https://www.sciencedirect.com/science/article/pii/S1748013223000439},
author = {Hongmei Zhang and Jingmei Tang and Bo Li and Bailing Li and Zucheng Zhang and Kun He and Shun Shi and Xiaohua Shen and Jialing Liu and Ziwei Huang and Di Wang and Wei Deng and Miaomiao Liu and Xinyun Zhou and Xidong Duan}
}

@article{Hattab11,
    author = {Hattab, H. and N’Diaye, A. T. and Wall, D. and Jnawali, G. and Coraux, J. and Busse, C. and van Gastel, R. and Poelsema, B. and Michely, T. and Meyer zu Heringdorf, F.-J. and Horn-von Hoegen, M.},
    title = {Growth temperature dependent graphene alignment on Ir(111)},
    journal = {Applied Physics Letters},
    volume = {98},
    number = {14},
    pages = {141903},
    year = {2011},
    month = {04},
    issn = {0003-6951},
    doi = {10.1063/1.3548546},
    url = {https://doi.org/10.1063/1.3548546}
}

@article{Son22,
title = {Thermodynamic stability reversal of iron sulfides at the nanoscale: Insights into the iron sulfide formation in low-temperature aqueous solution},
journal = {Geochimica et Cosmochimica Acta},
volume = {338},
pages = {220-228},
year = {2022},
issn = {0016-7037},
doi = {https://doi.org/10.1016/j.gca.2022.10.021},
url = {https://www.sciencedirect.com/science/article/pii/S0016703722005610},
author = {Sangbo Son and Sung Pil Hyun and Laurent Charlet and Kideok D. Kwon}
}

@article{Rickard07,
author = {Rickard, David and Luther, George W.},
title = {Chemistry of Iron Sulfides},
journal = {Chemical Reviews},
volume = {107},
number = {2},
pages = {514-562},
year = {2007},
doi = {10.1021/cr0503658},
URL = {https://doi.org/10.1021/cr0503658}
}

@article{Yang16,
  title = {Strong-coupling superconductivity revealed by scanning tunneling microscope in tetragonal FeS},
  author = {Yang, Xiong and Du, Zengyi and Du, Guan and Gu, Qiangqiang and Lin, Hai and Fang, Delong and Yang, Huan and Zhu, Xiyu and Wen, Hai-Hu},
  journal = {Physical Review B},
  volume = {94},
  issue = {2},
  pages = {024521},
  numpages = {6},
  year = {2016},
  month = {Jul},
  publisher = {American Physical Society},
  doi = {10.1103/PhysRevB.94.024521},
  url = {https://link.aps.org/doi/10.1103/PhysRevB.94.024521}
}

@article{Koteski17,
title = {First-principles calculations of tetragonal FeX (X=S, Se, Te): Magnetism, hyperfine-interaction, and bonding},
journal = {Journal of Magnetism and Magnetic Materials},
volume = {441},
pages = {769-775},
year = {2017},
issn = {0304-8853},
doi = {https://doi.org/10.1016/j.jmmm.2017.06.092},
url = {https://www.sciencedirect.com/science/article/pii/S0304885317308466},
author = {V. Koteski and V.N. Ivanovski and A. Umićević and J. Belošević-Čavor and D. Toprek and H.-E. Mahnke},
}

@software{fleurCode,
  author       = {Wortmann, Daniel and
                  Michalicek, Gregor and
                  Baadji, Nadjib and
                  Betzinger, Markus and
                  Bihlmayer, Gustav and
                  Bröder, Jens and
                  Burnus, Tobias and
                  Enkovaara, Jussi and
                  Freimuth, Frank and
                  Friedrich, Christoph and
                  Gerhorst, Christian-Roman and
                  Granberg Cauchi, Sabastian and
                  Grytsiuk, Uliana and
                  Hanke, Andrea and
                  Hanke, Jan-Philipp and
                  Heide, Marcus and
                  Heinze, Stefan and
                  Hilgers, Robin and
                  Janssen, Henning and
                  Klüppelberg, Daniel Aaaron and
                  Kovacik, Roman and
                  Kurz, Philipp and
                  Lezaic, Marjana and
                  Madsen, Georg K. H. and
                  Mokrousov, Yuriy and
                  Neukirchen, Alexander and
                  Redies, Matthias and
                  Rost, Stefan and
                  Schlipf, Martin and
                  Schindlmayr, Arno and
                  Winkelmann, Miriam and
                  Blügel, Stefan},
  title        = {FLEUR},
  month        = 5,
  year         = 2023,
  publisher    = {Zenodo},
  version      = {MaX-R8.0},
  doi          = {10.5281/zenodo.7576163},
  url          = {https://doi.org/10.5281/zenodo.7576163},
  howpublished = {Zenodo: 10.5281/zenodo.7576163}
}

@article{Krakauer1979,
  title = {Linearized augmented plane-wave method for the electronic band structure of thin films},
  author = {Krakauer, H. and Posternak, M. and Freeman, A. J.},
  journal = {Physical Review B},
  volume = {19},
  issue = {4},
  pages = {1706--1719},
  numpages = {0},
  year = {1979},
  month = {Feb},
  publisher = {American Physical Society},
  doi = {10.1103/PhysRevB.19.1706},
  url = {https://link.aps.org/doi/10.1103/PhysRevB.19.1706}
}

@article{Wimmer1981flapw,
  author = {Wimmer, E. and Krakauer, H. and Weinert, M. and Freeman, A. J.},
  title = {Full-potential self-consistent linearized-augmented-plane-wave method for calculating the electronic structure of mol\-e\-cules and sur\-faces: ${\mathrm{O}}_{2}$ mol\-e\-cule},
  journal = {Physical Review B},
  year = {1981},
  volume = {24},
  pages = {864--875},
  month = {7},
  doi = {10.1103/PhysRevB.24.864},
  number = {2},
  publisher = {American Physical Society}
}

@article{WeinertWimmerFreeman1982,
  title = {Total-energy all-electron density functional method for bulk solids and surfaces},
  author = {Weinert, M. and Wimmer, E. and Freeman, A. J.},
  journal = {Physical Review B},
  volume = {26},
  issue = {8},
  pages = {4571--4578},
  year = {1982},
  publisher = {American Physical Society},
  doi = {10.1103/PhysRevB.26.4571},
  url = {https://link.aps.org/doi/10.1103/PhysRevB.26.4571}
}

@article{PBE1996,
  author = {Perdew, John P. and Burke, Kieron and Ernzerhof, Matthias},
  title = {Generalized Gradient Approximation Made Simple},
  journal = {Physical Review Letters},
  volume = {77},
  pages = {3865--3868},
  year = {1996},
  doi = {10.1103/PhysRevLett.77.3865}
}

@article{Shimada98,
  title = {Spin-integrated and spin-resolved photoemission study of Fe chalcogenides},
  author = {Shimada, K. and Mizokawa, T. and Mamiya, K. and Saitoh, T. and Fujimori, A. and Ono, K. and Kakizaki, A. and Ishii, T. and Shirai, M. and Kamimura, T.},
  journal = {Physical Review B},
  volume = {57},
  issue = {15},
  pages = {8845--8853},
  numpages = {0},
  year = {1998},
  month = {Apr},
  publisher = {American Physical Society},
  doi = {10.1103/PhysRevB.57.8845},
  url = {https://link.aps.org/doi/10.1103/PhysRevB.57.8845}
}

@Article{Bansal20,
author={Bansal, Dipanshu
and Niedziela, Jennifer L.
and Calder, Stuart
and Lanigan-Atkins, Tyson
and Rawl, Ryan
and Said, Ayman H.
and Abernathy, Douglas L.
and Kolesnikov, Alexander I.
and Zhou, Haidong
and Delaire, Olivier},
title={Magnetically driven phonon instability enables the metal--insulator transition in h-FeS},
journal={Nature Physics},
year={2020},
month={Jun},
day={01},
volume={16},
number={6},
pages={669-675},
issn={1745-2481},
doi={10.1038/s41567-020-0857-1},
url={https://doi.org/10.1038/s41567-020-0857-1}
}

@article{Craco16,
    author = {Craco, L. and Faria, J. L. B.},
    title = {Electronic localization and bad-metallicity in pure and electron-doped troilite: A local-density-approximation plus dynamical-mean-field-theory study of FeS for lithium-ion batteries},
    journal = {Journal of Applied Physics},
    volume = {119},
    number = {8},
    pages = {085107},
    year = {2016},
    month = {02},
    issn = {0021-8979},
    doi = {10.1063/1.4942843},
    url = {https://doi.org/10.1063/1.4942843}
}

@Article{Fakhraee25,
author={Fakhraee, Mojtaba
and Crockford, Peter W.
and Bauer, Kohen W.
and Pasquier, Virgil
and Sugiyama, Ichiko
and Katsev, Sergei
and Raven, Morgan Reed
and Gomes, Maya
and Philippot, Pascal
and Crowe, Sean. A.
and Tarhan, Lidya G.
and Lyons, Timothy W.
and Planavsky, Noah},
title={The history of Earth's sulfur cycle},
journal={Nature Reviews Earth {\&} Environment},
year={2025},
month={Feb},
day={01},
volume={6},
number={2},
pages={106-125},
issn={2662-138X},
doi={10.1038/s43017-024-00615-0},
url={https://doi.org/10.1038/s43017-024-00615-0}
}

@Article{Walder05,
author={Walder, P.
and Pelton, A. D.},
title={Thermodynamic modeling of the Fe-S system},
journal={Journal of Phase Equilibria and Diffusion},
year={2005},
month={Feb},
day={01},
volume={26},
number={1},
pages={23-38},
issn={1863-7345},
doi={10.1007/s11669-005-0055-y},
url={https://doi.org/10.1007/s11669-005-0055-y}
}

@article{Evans70,
author = {Howard T. Evans},
title = {Lunar Troilite: Crystallography},
journal = {Science},
volume = {167},
number = {3918},
pages = {621-623},
year = {1970},
doi = {10.1126/science.167.3918.621},
URL = {https://www.science.org/doi/abs/10.1126/science.167.3918.621}
}

@article{Brett67,
title = {The occurrence and origin of lamellar troilite in iron meteorites},
journal = {Geochimica et Cosmochimica Acta},
volume = {31},
number = {5},
pages = {721-730},
year = {1967},
issn = {0016-7037},
doi = {https://doi.org/10.1016/S0016-7037(67)80027-9},
url = {https://www.sciencedirect.com/science/article/pii/S0016703767800279},
author = {Robin Brett and E.P. Henderson},
}

@article{Waechtershaeuser92,
title = {Groundworks for an evolutionary biochemistry: The iron-sulphur world},
journal = {Progress in Biophysics and Molecular Biology},
volume = {58},
number = {2},
pages = {85-201},
year = {1992},
issn = {0079-6107},
doi = {https://doi.org/10.1016/0079-6107(92)90022-X},
url = {https://www.sciencedirect.com/science/article/pii/007961079290022X},
author = {Günter Wächtershäuser}
}

@article{Martin03,
    author = {Martin, William and Russell, Michael J.},
    title = {On the origins of cells: a hypothesis for the evolutionary transitions from abiotic geochemistry to chemoautotrophic prokaryotes, and from prokaryotes to nucleated cells},
    journal = {Philosophical Transactions of the Royal Society B: Biological Sciences},
    volume = {358},
    number = {1429},
    pages = {59-85},
    year = {2003},
    month = {01},
    issn = {0962-8436},
    doi = {10.1098/rstb.2002.1183},
    url = {https://doi.org/10.1098/rstb.2002.1183}
}

@Article{Helmbrecht25,
author={Helmbrecht, Vanessa
and Reichelt, Robert
and Grohmann, Dina
and Orsi, William D.},
title={Simulated early Earth geochemistry fuels a hydrogen-dependent primordial metabolism},
journal={Nature Ecology {\&} Evolution},
year={2025},
month={May},
day={01},
volume={9},
number={5},
pages={769-778},
issn={2397-334X},
doi={10.1038/s41559-025-02676-w},
url={https://doi.org/10.1038/s41559-025-02676-w}
}

@Article{Jorgensen82,
author={J{\o}rgensen, Bo Barker},
title={Mineralization of organic matter in the sea bed---the role of sulphate reduction},
journal={Nature},
year={1982},
month={Apr},
day={01},
volume={296},
number={5858},
pages={643-645},
issn={1476-4687},
doi={10.1038/296643a0},
url={https://doi.org/10.1038/296643a0}
}

@article{Jorgensen19,
author = {Jørgensen, Bo Barker  and Findlay, Alyssa J.  and Pellerin, André},
title = {The Biogeochemical Sulfur Cycle of Marine Sediments},
journal = {Frontiers in Microbiology},
volume={10},
year={2019},
pages = {849},
doi={10.3389/fmicb.2019.00849},
url = {https://www.frontiersin.org/journals/microbiology/articles/10.3389/fmicb.2019.00849}
}

@article{Miao17,
author = {Miao, Ran and Dutta, Biswanath and Sahoo, Sanjubala and He, Junkai and Zhong, Wei and Cetegen, Shaylin A. and Jiang, Ting and Alpay, S. Pamir and Suib, Steven L.},
title = {Mesoporous Iron Sulfide for Highly Efficient Electrocatalytic Hydrogen Evolution},
journal = {Journal of the American Chemical Society},
volume = {139},
number = {39},
pages = {13604-13607},
year = {2017},
doi = {10.1021/jacs.7b07044},
URL = {https://doi.org/10.1021/jacs.7b07044},
}

@Article{Zou18,
author={Zou, Xiaoxin
and Wu, Yuanyuan
and Liu, Yipu
and Liu, Dapeng
and Li, Wang
and Gu, Lin
and Liu, Huan
and Wang, Pengwei
and Sun, Lei
and Zhang, Yu},
title={<em>In Situ</em> Generation of Bifunctional, Efficient Fe-Based Catalysts from Mackinawite Iron Sulfide for Water Splitting},
journal={Chem},
year={2018},
month={May},
day={10},
publisher={Elsevier},
volume={4},
number={5},
pages={1139-1152},
issn={2451-9294},
doi={10.1016/j.chempr.2018.02.023},
url={https://doi.org/10.1016/j.chempr.2018.02.023}
}

@article{Wang15,
author = {Wang, Di-Yan and Gong, Ming and Chou, Hung-Lung and Pan, Chun-Jern and Chen, Hsin-An and Wu, Yingpeng and Lin, Meng-Chang and Guan, Mingyun and Yang, Jiang and Chen, Chun-Wei and Wang, Yuh-Lin and Hwang, Bing-Joe and Chen, Chia-Chun and Dai, Hongjie},
title = {Highly Active and Stable Hybrid Catalyst of Cobalt-Doped FeS2 Nanosheets–Carbon Nanotubes for Hydrogen Evolution Reaction},
journal = {Journal of the American Chemical Society},
volume = {137},
number = {4},
pages = {1587-1592},
year = {2015},
doi = {10.1021/ja511572q},
URL = {https://doi.org/10.1021/ja511572q}
}

@article{Farhan24,
author = {Farhan, Ahmad and Qayyum, Wajeeha and Fatima, Urooj and Nawaz, Shahid and Balčiūnaitė, Aldona and Kim, Tak H. and Srivastava, Varsha and Vakros, John and Frontistis, Zacharias and Boczkaj, Grzegorz},
title = {Powering the Future by Iron Sulfide Type Material (FexSy) Based Electrochemical Materials for Water Splitting and Energy Storage Applications: A Review},
journal = {Small},
volume = {20},
number = {33},
pages = {2402015},
doi = {https://doi.org/10.1002/smll.202402015},
url = {https://onlinelibrary.wiley.com/doi/abs/10.1002/smll.202402015},
year = {2024}
}

@article{Gong16,
title = {Application of iron sulfide particles for groundwater and soil remediation: A review},
journal = {Water Research},
volume = {89},
pages = {309-320},
year = {2016},
issn = {0043-1354},
doi = {https://doi.org/10.1016/j.watres.2015.11.063},
url = {https://www.sciencedirect.com/science/article/pii/S0043135415303894},
author = {Yanyan Gong and Jingchun Tang and Dongye Zhao},
}

@article{Fan17,
author = {Fan, Dimin and Lan, Ying and Tratnyek, Paul G. and Johnson, Richard L. and Filip, Jan and O’Carroll, Denis M. and Nunez Garcia, Ariel and Agrawal, Abinash},
title = {Sulfidation of Iron-Based Materials: A Review of Processes and Implications for Water Treatment and Remediation},
journal = {Environmental Science \& Technology},
volume = {51},
number = {22},
pages = {13070-13085},
year = {2017},
doi = {10.1021/acs.est.7b04177},
URL = {https://doi.org/10.1021/acs.est.7b04177}
}

@article{Sano20,
  title={Structure changes of nanocrystalline mackinawite under hydrothermal conditions},
  author={Yoshinari Sano and Atsushi Kyono and Yasuhiro Yoneda and Noriko Isaka and Sota Takagi and Gen–ichiro Yamamoto},
  journal={Journal of Mineralogical and Petrological Sciences},
  volume={115},
  number={3},
  pages={261-275},
  year={2020},
  doi={10.2465/jmps.190903}
}

@article{Shigekawa19,
author = {Koshin Shigekawa  and Kosuke Nakayama  and Masato Kuno  and Giao N. Phan  and Kenta Owada  and Katsuaki Sugawara  and Takashi Takahashi  and Takafumi Sato },
title = {Dichotomy of superconductivity between monolayer FeS and FeSe},
journal = {Proceedings of the National Academy of Sciences},
volume = {116},
number = {49},
pages = {24470-24474},
year = {2019},
doi = {10.1073/pnas.1912836116},
URL = {https://www.pnas.org/doi/abs/10.1073/pnas.1912836116}
}

@article{Li96,
title = {Phase transitions in near stoichiometric iron sulfide},
journal = {Journal of Alloys and Compounds},
volume = {238},
number = {1},
pages = {73-80},
year = {1996},
issn = {0925-8388},
doi = {https://doi.org/10.1016/0925-8388(96)02207-4},
url = {https://www.sciencedirect.com/science/article/pii/0925838896022074},
author = {Fan Li and Hugo F. Franzen}
}

@Article{Takagi25,
author={Takagi, Rina
and Hirakida, Ryosuke
and Settai, Yuki
and Oiwa, Rikuto
and Takagi, Hirotaka
and Kitaori, Aki
and Yamauchi, Kensei
and Inoue, Hiroki
and Yamaura, Jun-ichi
and Nishio-Hamane, Daisuke
and Itoh, Shinichi
and Aji, Seno
and Saito, Hiraku
and Nakajima, Taro
and Nomoto, Takuya
and Arita, Ryotaro
and Seki, Shinichiro},
title={Spontaneous Hall effect induced by collinear antiferromagnetic order at room temperature},
journal={Nature Materials},
year={2025},
month={Jan},
day={01},
volume={24},
number={1},
pages={63-68},
issn={1476-4660},
doi={10.1038/s41563-024-02058-w},
url={https://doi.org/10.1038/s41563-024-02058-w}
}

@article{Huttmann15,
  title = {Tuning the van der Waals Interaction of Graphene with Molecules via Doping},
  author = {Huttmann, Felix and Mart\'{\i}nez-Galera, Antonio J. and Caciuc, Vasile and Atodiresei, Nicolae and Schumacher, Stefan and Standop, Sebastian and Hamada, Ikutaro and Wehling, Tim O. and Bl\"ugel, Stefan and Michely, Thomas},
  journal = {Phys. Rev. Lett.},
  volume = {115},
  issue = {23},
  pages = {236101},
  numpages = {6},
  year = {2015},
  month = {Dec},
  publisher = {American Physical Society},
  doi = {10.1103/PhysRevLett.115.236101},
  url = {https://link.aps.org/doi/10.1103/PhysRevLett.115.236101}
}

@article{Hohenberg64,
  title = {Inhomogeneous Electron Gas},
  author = {Hohenberg, P. and Kohn, W.},
  journal = {Phys. Rev.},
  volume = {136},
  issue = {3B},
  pages = {B864--B871},
  numpages = {0},
  year = {1964},
  month = {Nov},
  publisher = {American Physical Society},
  doi = {10.1103/PhysRev.136.B864},
  url = {https://link.aps.org/doi/10.1103/PhysRev.136.B864}
}

@Article{Hunger07,
author={Hunger, Stefan
and Benning, Liane G.},
title={Greigite: a true intermediate on the polysulfide pathway to pyrite},
journal={Geochemical Transactions},
year={2007},
month={Mar},
day={21},
volume={8},
number={1},
pages={1},
issn={1467-4866},
doi={10.1186/1467-4866-8-1},
url={https://doi.org/10.1186/1467-4866-8-1}
}

@article{Lan14,
title = {Monitoring the transformation of mackinawite to greigite and pyrite on polymer supports},
journal = {Applied Geochemistry},
volume = {50},
pages = {1-6},
year = {2014},
issn = {0883-2927},
doi = {https://doi.org/10.1016/j.apgeochem.2014.07.020},
url = {https://www.sciencedirect.com/science/article/pii/S0883292714001838},
author = {Ying Lan and Elizabeth C. Butler}
}

@article{Kobayashi01,
  title = {Pressure-induced semiconductor-metal-semiconductor transitions in FeS},
  author = {Kobayashi, Hisao and Takeshita, Nao and M\^ori, Nobuo and Takahashi, Hiroki and Kamimura, Takashi},
  journal = {Physical Review B},
  volume = {63},
  issue = {11},
  pages = {115203},
  numpages = {6},
  year = {2001},
  month = {Mar},
  publisher = {American Physical Society},
  doi = {10.1103/PhysRevB.63.115203},
  url = {https://link.aps.org/doi/10.1103/PhysRevB.63.115203}
}

@article{Sines12,
title = {Synthesis of tetragonal mackinawite-type FeS nanosheets by solvothermal crystallization},
journal = {Journal of Solid State Chemistry},
volume = {196},
pages = {17-20},
year = {2012},
issn = {0022-4596},
doi = {https://doi.org/10.1016/j.jssc.2012.07.056},
url = {https://www.sciencedirect.com/science/article/pii/S0022459612004902},
author = {Ian T. Sines and Dimitri D. {Vaughn II} and Rajiv Misra and Eric J. Popczun and Raymond E. Schaak},
}

@article{Ersoy2011,
  title = {Effective Coulomb interaction in transition metals from constrained random-phase approximation},
  author = {\ifmmode \mbox{\c{S}}\else \c{S}\fi{}a\ifmmode \mbox{\c{s}}\else \c{s}\fi{}\ifmmode \imath \else \i \fi{}o\ifmmode \breve{g}\else \u{g}\fi{}lu, Ersoy and Friedrich, Christoph and Bl\"ugel, Stefan},
  journal = {Physical Review B},
  volume = {83},
  issue = {12},
  pages = {121101},
  numpages = {4},
  year = {2011},
  month = {Mar},
  publisher = {American Physical Society},
  doi = {10.1103/PhysRevB.83.121101},
  url = {https://link.aps.org/doi/10.1103/PhysRevB.83.121101}
}

@article{Prabhu2025,
author = {Prabhu, Mahesh Krishna and David, Philippe and Guisset, Val{\'e}rie and Martinelli, Lucio and Coraux, Johann and Renaud, Gilles},
title = {Reactive Molecular Beam Epitaxy Growth of a 1T-FeS2 Single-Layer–Atomic Structure, Moiré, and Decoupling via Intercalation},
journal = {ACS Nano},
volume = {19},
number = {14},
pages = {13941-13951},
year = {2025},
doi = {10.1021/acsnano.4c17873},
URL = {https://doi.org/10.1021/acsnano.4c17873}
}

@Article{Zhou2023,
author = {Zhou, Jiadong
and Zhu, Chao
and Zhou, Yao
and Dong, Jichen
and Li, Peiling
and Zhang, Zhaowei
and Wang, Zhen
and Lin, Yung-Chang
and Shi, Jia
and Zhang, Runwu
and Zheng, Yanzhen
and Yu, Huimei
and Tang, Bijun
and Liu, Fucai
and Wang, Lin
and Liu, Liwei
and Liu, Gui-Bin
and Hu, Weida
and Gao, Yanfeng
and Yang, Haitao
and Gao, Weibo
and Lu, Li
and Wang, Yeliang
and Suenaga, Kazu
and Liu, Guangtong
and Ding, Feng
and Yao, Yugui
and Liu, Zheng},
title = {Composition and phase engineering of metal chalcogenides and phosphorous chalcogenides},
journal = {Nature Materials},
year = {2023},
month = {Apr},
day = {01},
volume = {22},
number = {4},
pages = {450-458},
issn = {1476-4660},
doi = {10.1038/s41563-022-01291-5},
url = {https://doi.org/10.1038/s41563-022-01291-5}
}

@article{Lee10,
  title = {Higher-accuracy van der Waals density functional},
  author = {Lee, Kyuho and Murray, \'Eamonn D. and Kong, Lingzhu and Lundqvist, Bengt I. and Langreth, David C.},
  journal = {Physical Review B},
  volume = {82},
  issue = {8},
  pages = {081101},
  numpages = {4},
  year = {2010},
  month = {Aug},
  publisher = {American Physical Society},
  doi = {10.1103/PhysRevB.82.081101},
  url = {https://link.aps.org/doi/10.1103/PhysRevB.82.081101}
}

@article{Li26,
author = {Deng, Li and Wang, Fei and Yin, Xiang and Tong, Junwei and Wu, Yanzhao and Zhang, Xianmin},
title = {Perpendicular Néel Vector Detection of Monolayer Collinear Antiferromagnets with Spin-Layer Coupling},
journal = {Journal of the American Chemical Society},
volume = {148},
number = {1},
pages = {1837-1846},
year = {2026},
doi = {10.1021/jacs.5c19267},
URL = {https://doi.org/10.1021/jacs.5c19267}
}

@article{Long25,
  title = {Anomalous Hall effect driven by electric field and sliding ferroelectricity in two-dimensional compensated antiferromagnetic MnS},
  author = {Long, Guozhi and Huang, Huaqing},
  journal = {Physical Review B},
  volume = {112},
  issue = {16},
  pages = {165136},
  numpages = {7},
  year = {2025},
  month = {Oct},
  publisher = {American Physical Society},
  doi = {10.1103/z69z-175r},
  url = {https://link.aps.org/doi/10.1103/z69z-175r}
}

@article{Qayyum24,
author = {Qayyum, Hafiz Adil and Mansha, Muhammad and Sattar, Shahid},
title = {PT-Symmetry Breaking and Spin Control in 2D Antiferromagnetic MnSe},
journal = {ACS Omega},
volume = {9},
number = {47},
pages = {47097-47104},
year = {2024},
doi = {10.1021/acsomega.4c07291},
URL = {https://doi.org/10.1021/acsomega.4c07291}
}

@article{Wang26,
title = {Topological edge states orchestrate bifunctional HER/OER on FeS nanoribbons},
journal = {International Journal of Hydrogen Energy},
volume = {220},
pages = {154094},
year = {2026},
issn = {0360-3199},
doi = {https://doi.org/10.1016/j.ijhydene.2026.154094},
url = {https://www.sciencedirect.com/science/article/pii/S0360319926007317},
author = {Juan Wang and Yiming Lu and Xikui Ma and Hongxia Bu and Kehan Liu and Yueheng Du and Junru Wang and Han Gao and Yingcai Fan and Mingwen Zhao}
}

@Article{Liu2024,
author={Liu, Kehan
and Ma, Xikui
and Li, Yangyang
and Zhao, Mingwen},
title={Tailoring the quantum anomalous layer Hall effect in multiferroic bilayers through sliding},
journal={npj Computational Materials},
year={2024},
month={Jun},
day={03},
volume={10},
number={1},
pages={118},
issn={2057-3960},
doi={10.1038/s41524-024-01306-6},
url={https://doi.org/10.1038/s41524-024-01306-6}
}

@article{Liang24,
    author = {Liang, Yan and Han, Xuening and Wang, Qiang and Zhao, Pei},
    title = {Thickness-dependent topological phases in topological magnet Fe2S2},
    journal = {Applied Physics Letters},
    volume = {124},
    number = {6},
    pages = {062411},
    year = {2024},
    month = {02},
    issn = {0003-6951},
    doi = {10.1063/5.0190298},
    url = {https://doi.org/10.1063/5.0190298}
}

@Article{Schmitt2019,
author={Schmitt, Martin
and Moras, Paolo
and Bihlmayer, Gustav
and Cotsakis, Ryan
and Vogt, Matthias
and Kemmer, Jeannette
and Belabbes, Abderrezak
and Sheverdyaeva, Polina M.
and Kundu, Asish K.
and Carbone, Carlo
and Bl{\"u}gel, Stefan
and Bode, Matthias},
title={Indirect chiral magnetic exchange through Dzyaloshinskii--Moriya-enhanced RKKY interactions in manganese oxide chains on Ir(100)},
journal={Nature Communications},
year={2019},
month={Jun},
day={13},
volume={10},
number={1},
pages={2610},
issn={2041-1723},
doi={10.1038/s41467-019-10515-3},
url={https://doi.org/10.1038/s41467-019-10515-3}
}

@article{Goikoetxea2022,
  title = {Multiplet effects in the electronic correlation of one-dimensional magnetic transition metal oxides on metals},
  author = {Goikoetxea, J. and Friedrich, C. and Bihlmayer, G. and Bl\"ugel, S. and Arnau, A. and Blanco-Rey, M.},
  journal = {Physical Review B},
  volume = {106},
  issue = {3},
  pages = {035130},
  numpages = {9},
  year = {2022},
  month = {Jul},
  publisher = {American Physical Society},
  doi = {10.1103/PhysRevB.106.035130},
  url = {https://link.aps.org/doi/10.1103/PhysRevB.106.035130}
}

@article{Keen1995,
doi = {10.1088/0953-8984/7/29/007},
url = {https://doi.org/10.1088/0953-8984/7/29/007},
year = {1995},
month = {jul},
publisher = {},
volume = {7},
number = {29},
pages = {5793},
author = {Keen, D. A. and Hull, S.},
title = {The high-temperature structural behaviour of copper(I) iodide},
journal = {Journal of Physics: Condensed Matter}
}

\clearpage
\end{document}

\DeclareGraphicsExtensions{.pdf}

\title{Supporting Information:\\
Emergence of a non-bulk hexagonal Fe$_2$S$_2$ single layer via phase transformation}

\author{Affan Safeer}
\affiliation{II. Physikalisches Institut, Universit\"{a}t zu K\"{o}ln, Z\"{u}lpicher Str. 77, 50937 Cologne, Germany \looseness=-1}
\author{Wejdan Beida}
\affiliation{Peter Gr\"{u}nberg Institut, Forschungszentrum J\"{u}lich, 52425 J\"{u}lich, Germany} 
\affiliation{Institute for Theoretical Physics, RWTH Aachen University, 52062 Aachen, Germany}
\author{Felix Oberbauer}
\affiliation{II. Physikalisches Institut, Universit\"{a}t zu K\"{o}ln, Z\"{u}lpicher Str. 77, 50937 Cologne, Germany \looseness=-1}
\author{Nicolae Atodiresei}
\affiliation{Peter Gr\"{u}nberg Institut, Forschungszentrum J\"{u}lich, 52425 J\"{u}lich, Germany} 
\author{Gustav Bihlmayer}
\affiliation{Peter Gr\"{u}nberg Institut, Forschungszentrum J\"{u}lich, 52425 J\"{u}lich, Germany} 
\author{Max Wolfertz}
\affiliation{II. Physikalisches Institut, Universit\"{a}t zu K\"{o}ln, Z\"{u}lpicher Str. 77, 50937 Cologne, Germany \looseness=-1}
\author{Chiara Schlichte}
\affiliation{II. Physikalisches Institut, Universit\"{a}t zu K\"{o}ln, Z\"{u}lpicher Str. 77, 50937 Cologne, Germany \looseness=-1}
\author{Wouter Jolie}
\affiliation{II. Physikalisches Institut, Universit\"{a}t zu K\"{o}ln, Z\"{u}lpicher Str. 77, 50937 Cologne, Germany \looseness=-1}
\author{Stefan Blügel}
\affiliation{Peter Gr\"{u}nberg Institut, Forschungszentrum J\"{u}lich, 52425 J\"{u}lich, Germany} 
\affiliation{Institute for Theoretical Physics, RWTH Aachen University, 52062 Aachen, Germany}
\author{Jeison Fischer}
\email{jfischer@ph2.uni-koeln.de}
\affiliation{II. Physikalisches Institut, Universit\"{a}t zu K\"{o}ln, Z\"{u}lpicher Str. 77, 50937 Cologne, Germany \looseness=-1}
\author{Thomas Michely}
\affiliation{II. Physikalisches Institut, Universit\"{a}t zu K\"{o}ln, Z\"{u}lpicher Str. 77, 50937 Cologne, Germany \looseness=-1}
%\date{\today}

\maketitle
%\newpage
%\tableofcontents

%\newpage
%\subsection*{Supplementary Note 1: Comparison of experimental observations with single-layer Fe--S polymorphs from \href{https://c2db.fysik.dtu.dk/}{C2DB} database.}
%\begin{figure}[hbt!]
%\includegraphics[width=\textwidth]{move.pdf}
%\caption{Table S1. Single-layer FeS polymorph from C2DB database \cite{} \tm{Comment TM: Dear Affan, please insert the table from the magnetism SI.

\newpage
\subsection*{Supplementary Note 1. $\mathrm{d}I/\mathrm{d}V$ point spectrum on bilayer island compared $\mathrm{d}I/\mathrm{d}V$ point spectrum on single-layer island of h-Fe$_2$S$_2$}
\begin{figure}[hbt!]
\includegraphics[width=\textwidth]{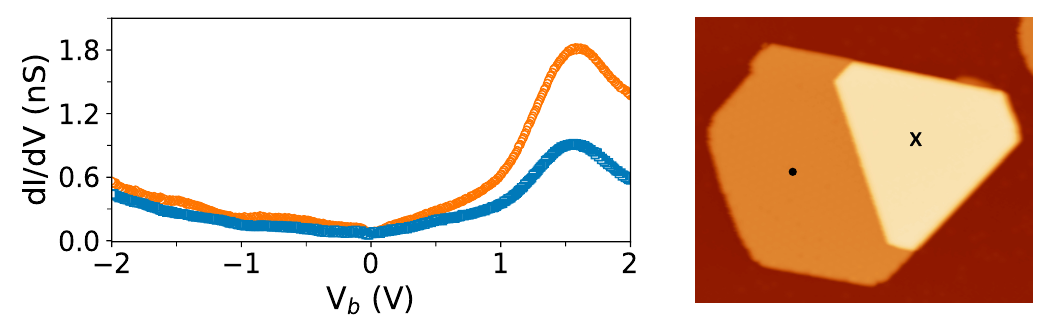}
\caption{$\mathrm{d}I/\mathrm{d}V$ point spectra of single-layer (orange curve) and bilayer (blue curve) of h-Fe$_2$S$_2$ ($V_{st} =2$\,V, $I_{st}=2$\,nA, $f_{mod} =677$\,Hz, and $V_{mod}= 20$\,mV). The spectra are taken at the locations marked $\bullet$ and \textbf{x} on single layer and bilayer h-Fe$_2$S$_2$ as shown in the STM image ($V_\mathrm{b} = 1.8$\,V, $I_\mathrm{t} = 50$\,pA, 90\,nm $\times$ 80\,nm).
}
\label{} 
\end{figure}

\newpage
\subsection*{Supplementary Note 2: Comparison of experimental observations with single-layer FeS polymorphs from C2DB database.}

Table S1 contains the t-Fe$_2$S$_2$ structure (single layer mackinawite) as first entry and all single layer hexagonal FeS polymorphs obtained from high-throughput calculations in the C2DB \cite{Haastrup_2018,Gjerding_2021} database. From the polymorphs presented, entry 1: Fe$_2$S$_2$ \cite{Zhao2017,Lin2018,Shigekawa19}, entry 4: FeS$_2$ \cite{Prabhu2025}, and entry 5: Fe$_3$S$_4$ \cite{Lin2018} were experimentally reported as single layer materials.

\clearpage
\begin{table}[htbp]
\centering
\caption{Single-layer FeS polymorph from C2DB database. The magnetic ground states NM, M, and N/A refer to non-magnetic, magnetic, and not known, respectively.}
\label{FeS_structures}
\begin{tabular}{|l|l|c|c|c|c|c|}
\hline
stoichiometry & structure view & Fe & lattice $a$ (\AA) & magnetic & form. E (eV/atom) & deviation exp. \\ \hline

1. Fe$_2$S$_2$ (p4/nmm) & \includegraphics[width=30mm]{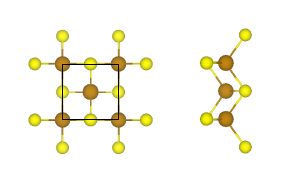} & 2 & 3.599 & NM & -0.496 & -2\% \\ \hline

2. FeS$_2$ (p2$_1$/m) & \includegraphics[width=30mm]{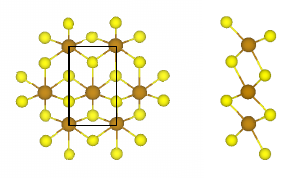} & 1 & 3.227 & NM & -0.383 & -14\% \\ \hline

3. Fe$_2$S$_2$ (p$\bar{3}$m1) & \includegraphics[width=30mm]{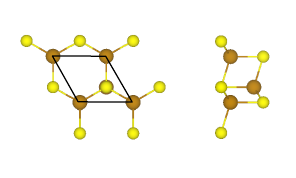} & 2 & 3.574 & N/A & -0.358 & -5\% \\ \hline

4. FeS$_2$ (p$\bar{3}$m1) & \includegraphics[width=30mm]{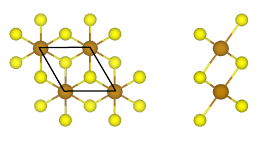} & 1 & 3.203 & M & -0.326 & -15\% \\ \hline

5. Fe$_3$S$_4$ (p$\bar{3}$m1) & \includegraphics[width=30mm]{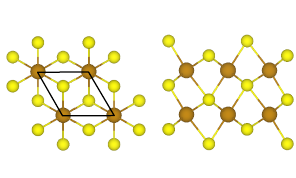} & 3 & 3.307 & M & -0.309 & -12\% \\ \hline

6. FeS$_2$ (p$\bar{6}$m2) & \includegraphics[width=30mm]{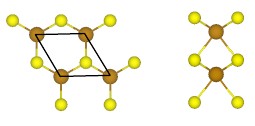} & 1 & 3.154 & M & -0.263 & -16\% \\ \hline

7. Fe$_3$S$_4$ (p$\bar{6}$m2) & \includegraphics[width=30mm]{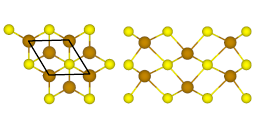} & 3 & 3.077 & M & -0.177 & -18\% \\ \hline

8. Fe$_2$S$_2$ (p$\bar{3}$m1) & \includegraphics[width=30mm]{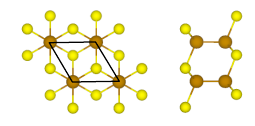} & 2 & 3.393 & NM & -0.106 & -10\% \\ \hline

9. Fe$_2$S$_2$ (p$\bar{3}$m1) & \includegraphics[width=30mm]{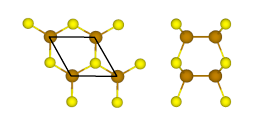} & 2 & 3.418 & M & -0.094 & -9\% \\ \hline

\end{tabular}
\end{table}

\clearpage
\subsection*{Supplementary Note 3. DFT calculation of interlayer spacing for  h-Fe$_2$S$_2$}
\begin{figure}[hbt!]
\includegraphics[width=0.5 \textwidth]{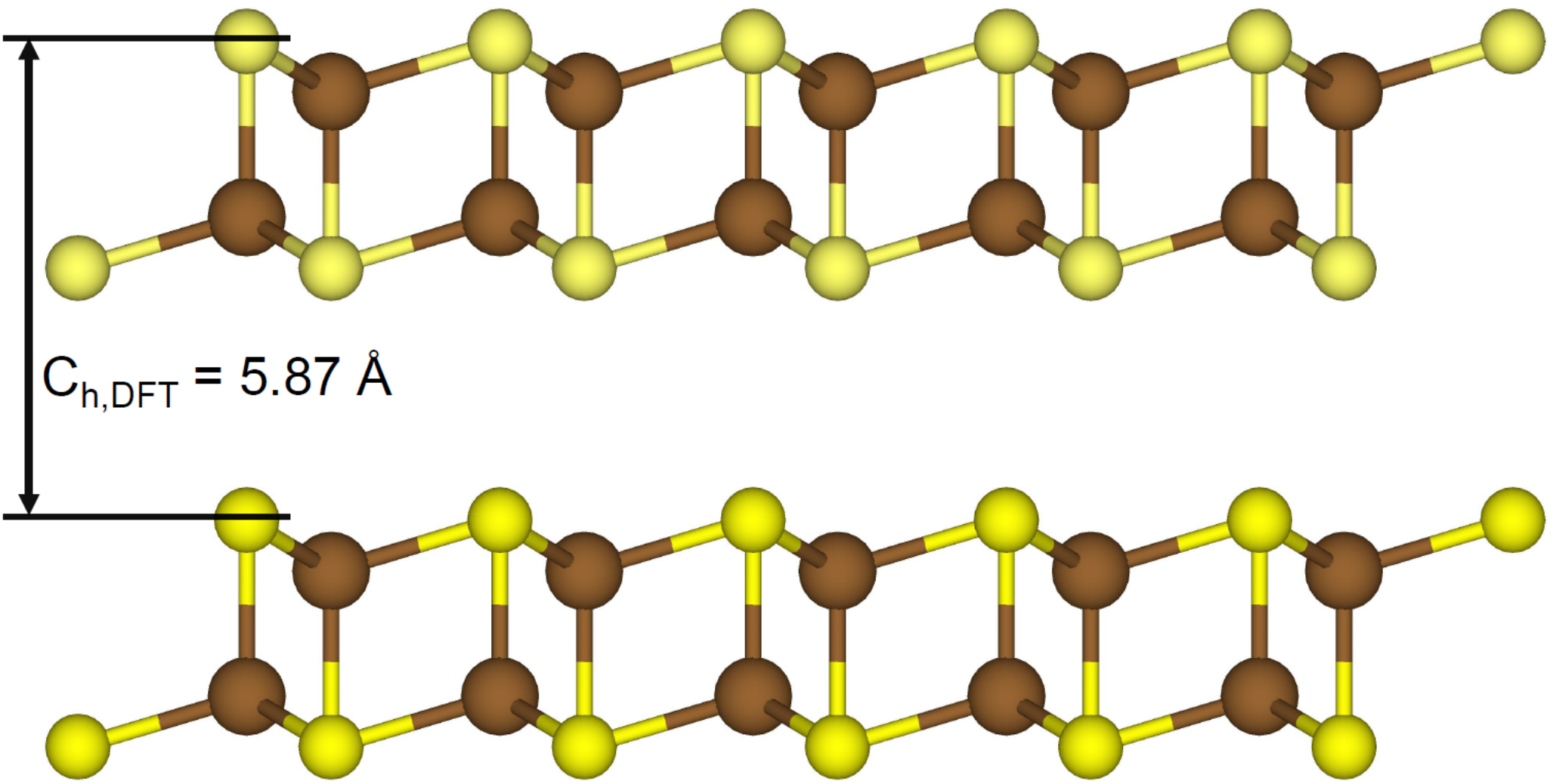}
\caption{Side-view ball model of fully relaxed non-spinpolarized DFT calculation for a bilayer h-Fe$_2$S$_2$ with an in-plane lattice constant of 3.56~\AA. Including the van der Waals interactions into the relaxations we obtained an interlayer distance of $c_\mathrm{h,DFT} = 5.87$\,\AA.
}
\label{SI_c distance} 
\end{figure}

 The \textit{ab intio} calculations for Figure~S2 were performed using DFT \cite{Hohenberg64} with the projector augmented plane wave method \cite{PRB50_17953} as implemented in the VASP code \cite{PRB47_558, PRB54_11169}. A cutoff energy of 500\,eV 
 was used for the plane wave expansion of the Kohn-Sham wave 
 functions \cite{PR140_A1133}. 
We modeled the bilayer by using the supercell approach with a vertical lattice constant of 25~\AA\ that includes a sufficiently large vacuum spacing to eliminate spurious interactions between periodic images. 
The structural relaxations were performed using the vdW-DF2 functional \cite{Lee10} with a revised Becke (B86b) exchange \cite{JCP86_7184,PRB89_121103R}, a generalized-gradient-approximated (GGA) functional to properly account for the nonlocal correlation effects like van der Waals interactions \cite{Huttmann15}.

\newpage
\subsection*{Supplementary Note 4. Magnetic spin configurations in h-Fe$_2$S$_2$}
\begin{figure}[hbt!]
\includegraphics[width=0.9\textwidth]{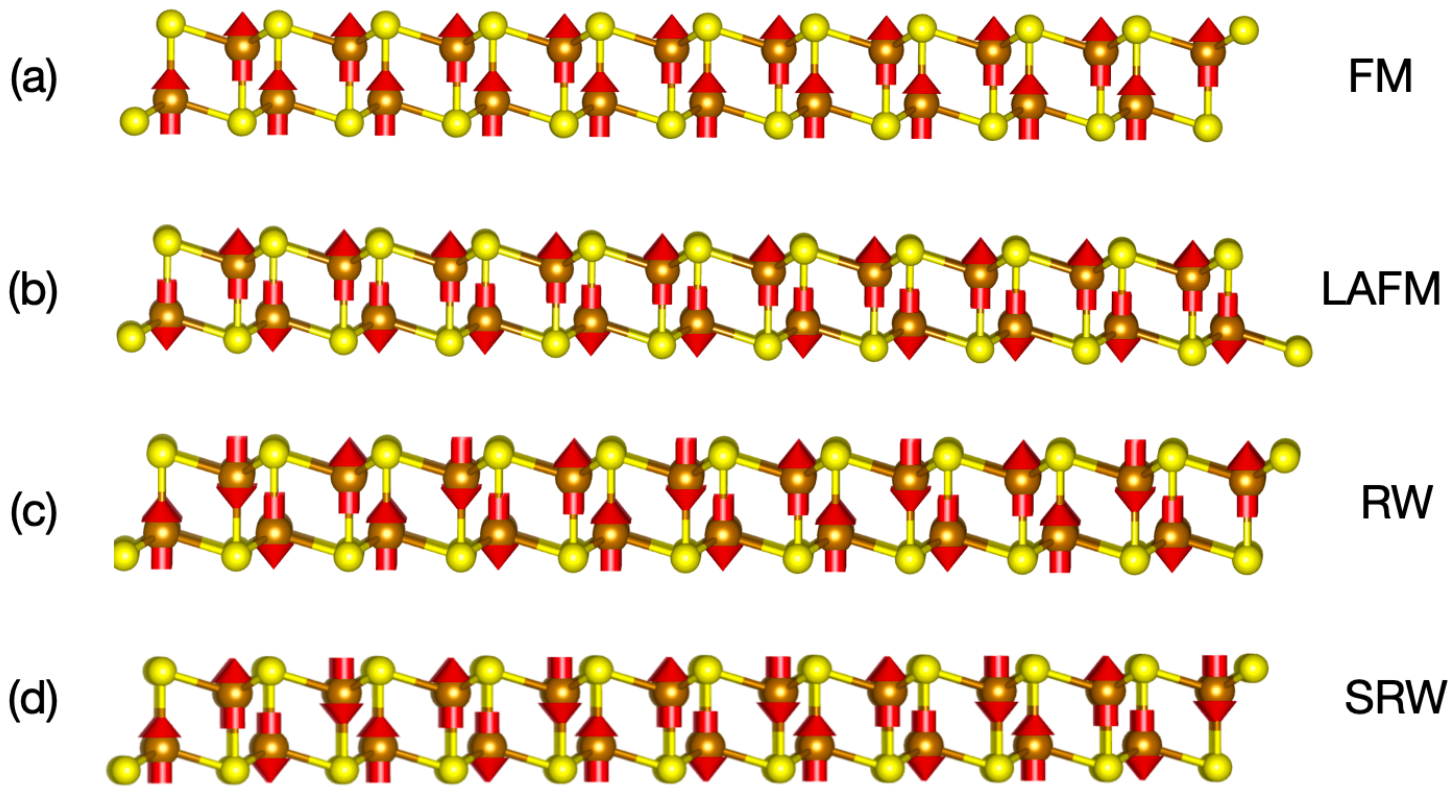}
\caption{Side view of the monolayer of hexagonal Fe$_2$S$_2$ illustrating the vertical stacking of FeS sublayers that form a buckled bilayer honeycomb structure. Different magnetic configurations considered in this work are shown:
(a) ferromagnetic (FM), (b) layered antiferromagnetic (LAFM), (c) row-wise antiferromagnetic (RW), and (d) shifted row-wise antiferromagnetic (SRW). RW and SRW differ by a phase shift of the spin pattern of adjacent sublayers, with SRW corresponding to a one-site translation of the RW  order. Yellow and brown spheres represent S and Fe atoms, respectively, while arrows indicate the orientation of local magnetic moments.}
\label{SI_hFeS_spin} 
\end{figure}

\newpage
\subsection*{Supplementary Note 5. DFT calculations of t-Fe$_2$S$_2$ and h-Fe$_2$S$_2$ clusters on graphene.}
\begin{figure}[hbt!]
\includegraphics[width=\textwidth]{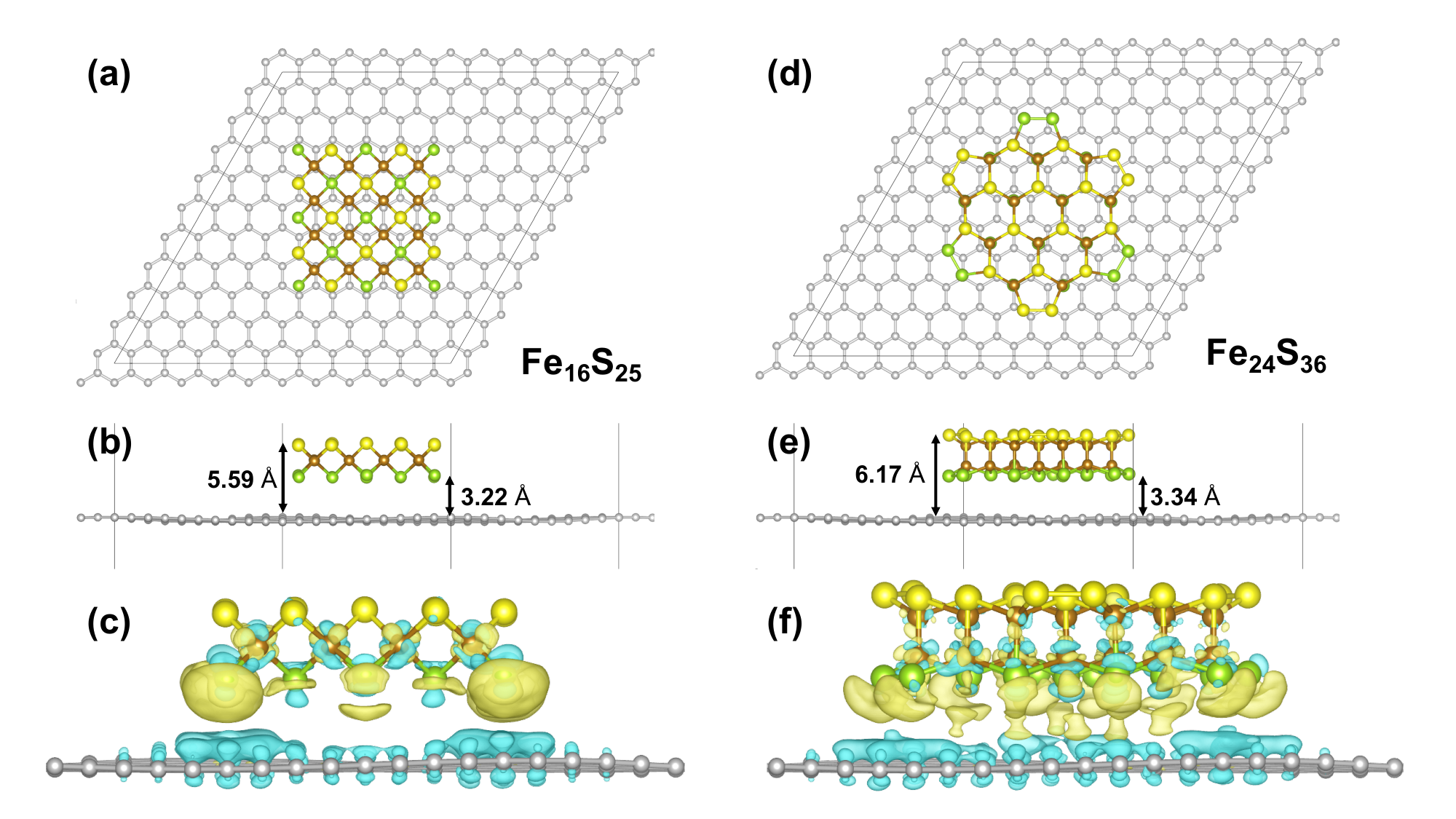}
\caption{(a) Top view and (b) side view ball model of DFT supercell for the t-Fe$_{16}$S$_{25}$ cluster on Gr. (c) Charge density difference plot representing isosurfaces ($1.29\times10^{-8}$ electrons/\AA$^3$) of charge accumulation (light yellow) and depletion (light blue) for the t-Fe$_{16}$S$_{25}$ cluster on Gr. (d)-(f) Same as (a)-(c), but for  h-Fe$_{24}$S$_{36}$. Brown balls: Fe atoms; yellow balls: upper layer S; green balls: bottom layer S. 
}
\label{SI_DFT} 
\end{figure}
The non-spinpolarized DFT calculations were conducted using methods as outlined in Supplementary Note 3. Clusters of t-Fe$_2$S$_2$ and h-Fe$_2$S$_2$ were adsorbed to a $(10\times10)$ Gr unit cell large enough to avoid cluster--cluster interaction of neighboring unit cells. The geometry of Gr corresponds to the one as relaxed on Ir(111) \cite{Busse11}. Only the cluster atoms were relaxed while the C atoms were fixed. The clusters possess S-terminated edges corresponding to the experimental growth in S excess. As obvious from the large distances $ > 3.2$\,\AA~of the clusters from the Gr substrate, the clusters bind to Gr predominantly via van der Waals forces. The charge density differences plotted in Figure~S3(c) and S3(f) represent the charge density rearrangement upon bonding. They indicate a weak charge transfer from Gr towards the S atoms of the bottom S layers and thus an additional weak electrostatic contribution to cluster binding. 
Since the clusters have different numbers of atoms and different in-plane surface areas, it is adequate to analyze their binding energies in terms of  meV/\AA$^2$.
For the adsorbed t-Fe$_{16}$S$_{25}$ cluster we find an interaction energy of $-27.17$ meV/\AA$^2$, for the adsorbed h-Fe$_{24}$S$_{36}$ cluster the interaction energy amounts to $-23.27$ meV/\AA$^2$. Although binding is affected by the finite size of the clusters, the interaction energies are consistent with van der Waals binding. For comparison, for Gr adsorbed on Ir(111) this energy is $-18.97$  meV/\AA$^2$. 

\newpage
\subsection*{Supplementary Note 6. Variations of properties with Hubbard $U$}
\begin{table}[hbt!]
\centering
\caption{Lattice constant $a$ in \AA\ and energy difference $E_{\text{RW}}-E_{\text{c(2$\times$2) AFM}}$ in meV between the ground state energies of  c(2$\times$2) antiferromagnetic  t-Fe$_2$S$_2$ and row-wise (RW) antiferromagnetic (AFM) h-Fe$_2$S$_2$ phases as a function of the Hubbard parameter $U$ in combination with the PBE functional for a given interorbital $J=0.63$~eV applied to Fe $d$ orbitals.  The sign-change in the energy difference indicates a structural phase transition of  the ground-state phase. For $U>1.4$\,eV ($<1.6$~eV) the RW-AFM-h (c(2$\times$2) AFM-t) phase has lower energy.}

\begin{tabular}{|c|c|c|c|} 
\hline
$U$ on $d$-Fe & $a$(c(2$\times$2) AFM t-Fe$_2$S$_2$)  & $a$(\text{RW h-Fe$_2$S$_2$}) & $E_{\text{RW}}-E_{\text{c(2$\times$2) AFM}}$ \\
 (eV)  & (\AA)   &  (\AA) &  (meV) \\
\hline
0       &  3.59 &  3.69  & \phantom{0}225.0\\ 
1.4     &  3.66 &  3.71  & \phantom{00}12.0 \\ 
1.6     &  3.71 &  3.72  & \phantom{0}$-$30.0\\ 
2.8     &  3.74 &  3.82  & $-$268.5 \\ 
\hline
\end{tabular}
\end{table}

\newpage
\subsection*{Supplementary Note 7. Spin-, orbital and atom-resolved electronic structure of c($2\times 2$) AFM t-Fe$_2$S$_2$}
\begin{figure}[hbt!]
\includegraphics[width=1.0\textwidth]{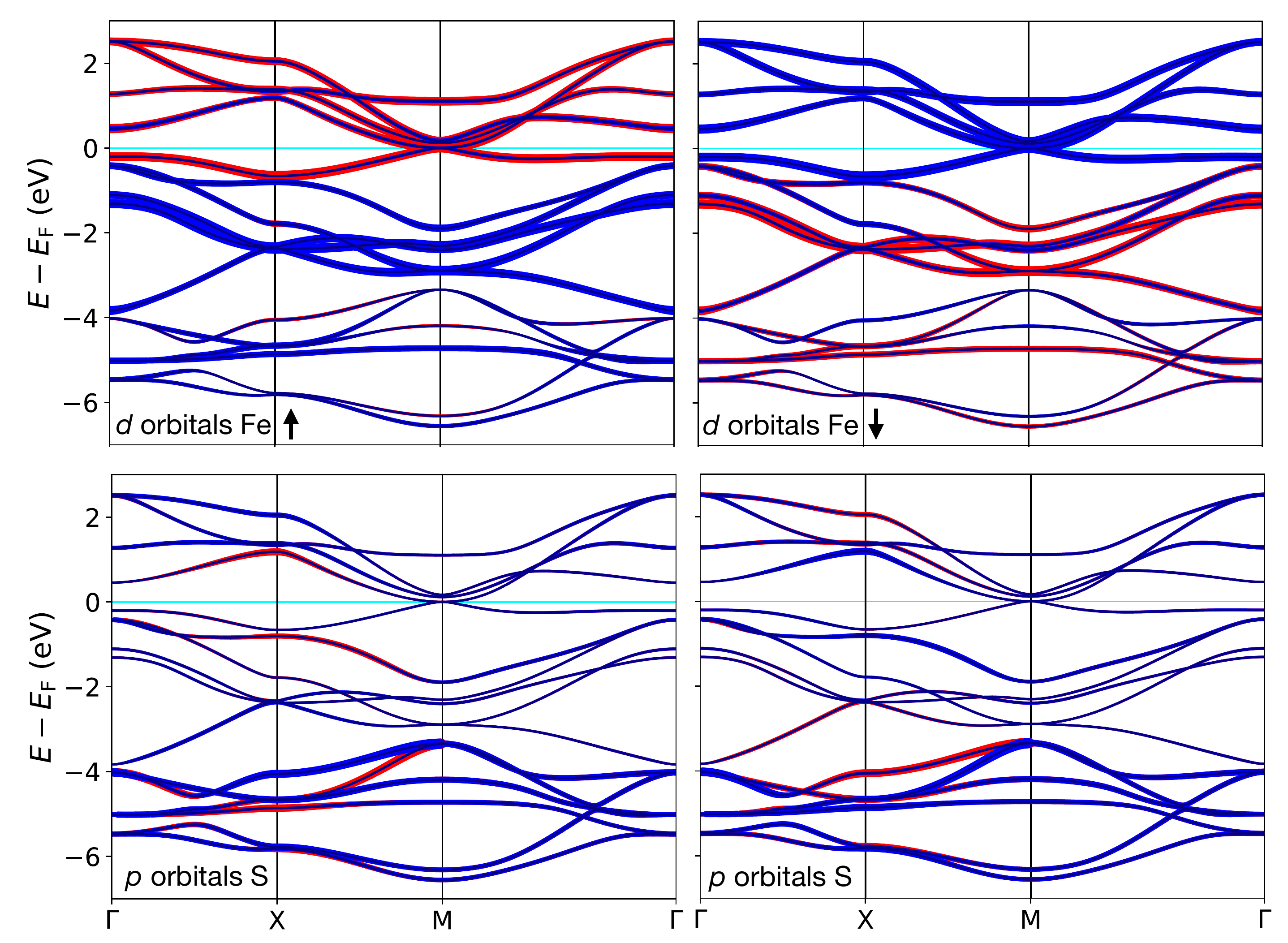}
\caption{Band structure of t-Fe$_2$S$_2$ in the c(2$\times$2) checkerboard AFM order from PBE+$U$, $U=1.6$~eV, $J=0.63$~eV, along the high-symmetry lines $\overline{\Gamma}$–$\overline{\mathrm{X}}$–$\overline{\mathrm{M}}$–$\overline{\Gamma}$  of the square Brillouin zone projected onto Fe-$d$ states (top panels)  and S-$p$ states (bottom panels) of Fe atoms with majority spins aligned in  $\uparrow$ and  $\downarrow$ projection,  as well as sulfur atoms in the upper and lower plane of Fe (left and right panels). Red and blue color denote projections into opposite spin channels. Blue (red) refers to spin-projections according to  $\uparrow$ ($\downarrow$) alignment.  Thickness of lines scales with the weight of projection. Energies are given relative to Fermi energy $E_\textrm{F}$.}
\label{SI_BS_t-Fe2S2_spin+atom_resolved}
\end{figure}

\begin{figure}[hbt!]
\includegraphics[width=\textwidth]{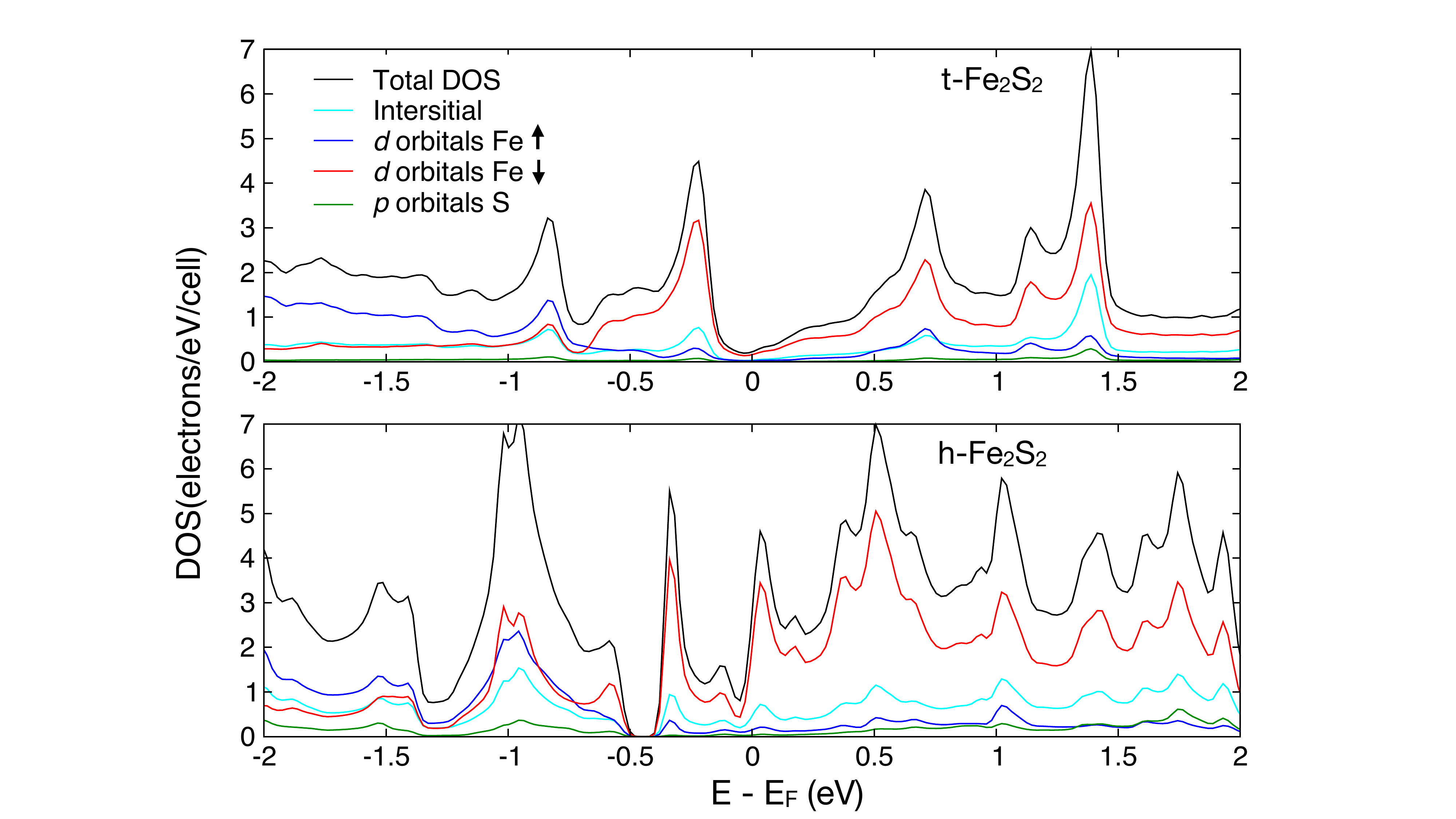}
\caption{Spin- and orbital-projected density of states (PDOS) for the monolayers of t-Fe$_2$S$_2$ in the c($2\times2$) AFM phase and of h-Fe$_2$S$_2$ in the RW AFM  structure calculated at the level of PBE$+U$ with  $U$=1.6~eV and $J=0.63$~eV applied to the $d$-orbitals of the Fe atoms. Shown are the spin-projections with respect to the $\uparrow$ spin-quantization axis. The total spin-projected density of states (DOS) is shown along with the contributions from Fe $d$ orbitals, S $p$ orbitals, both within the muffin-tin spheres of the respective atoms and the interstitial region between Fe and S atoms calculated in the entire unit cell.  The projection with respect to the $\downarrow$ spin-quantization axis is not shown, but is identical except, spin-up and -down stated are reversed. The total DOS is the sum of both projections.  The energy is referenced with respect to the Fermi energy $E_\text{F}$. Since the antiferromagnetically aligned Fe atoms are chemically identical, the red line indicates the minority-spin \textit{i.e.}\ spin-down $d$ orbitals of the Fe atom with up magnetization, and the blue line the corresponding majority (spin-up) electrons. Minority and Majority state change their spin-polarization for Fe atom with down-magnetization.}
\label{SI_pdos}
\end{figure}

Figure~\ref{SI_BS_t-Fe2S2_spin+atom_resolved}, the band structure of t-Fe$_2$S$_2$ in the c(2$\times$2) antiferromagnetic configuration reveals a clear orbital- and spin-dependent character. The states near the Fermi level are predominantly of Fe-$d$ character and exhibit a pronounced spin polarization, while the S-$p$ states are mainly located at lower binding energies and display a weaker, yet finite, spin polarization due to hybridization with Fe-$d$ orbitals. Importantly, a comparison between the chemically equivalent Fe atoms of spin-up and -down alignment shows an exact reversal of the spin-resolved spectral weight: bands that are majority-spin (blue) on Fe($\uparrow$) appear as majority-spin (red) on Fe($\downarrow$), and vice versa for minority-spin (red) on Fe($\uparrow$).  The same behavior is reflected in the S-$p$ projections, indicating that the induced spin polarization on sulfur follows the local exchange field of the neighboring Fe atoms. This systematic color inversion directly visualizes the N\'eel-type antiferromagnetic order, in which the electronic structure remains globally spin-degenerate while being locally spin-polarized on each sublattice.

This discussion can also be followed from the spin- and orbital-projected density of state (DOS) in Figure~\ref{SI_pdos}, which in addition to the t-phase shows also the analysis of the RW AFM h-Fe$_2$S$_2$ phase. Both polymorphs exhibit a pronounced spin-dependent electronic structure dominated by Fe  $d$-states in the vicinity of the Fermi level. The total spin-projected density of states (black curve) is largely composed of Fe $d$-orbital contributions, while S $p$-states (green) are primarily located at lower energies and contribute weakly near $E_\text{F}$.The interstitial contribution (cyan) is finite but remains smaller than the Fe $d$-weight, indicating that the electronic states are predominantly localized on the Fe sites. The t-phase exhibits a clear suppression of spectral weight at the Fermi level, consistent with the semi-metallic electronic structure. The low-energy electronic states are strongly spin-polarized:
The majority-spin Fe $d$-states (blue) are mostly located below 
$E_\text{F}$, forming well-defined peaks in the valence region.
The minority-spin Fe $d$-states (red) dominate the conduction region above $E_\text{F}$. This separation indicates a substantial exchange splitting of Fe $d$-orbitals, characteristic of localized magnetic moments in an AFM background. The relatively small S $p$-contribution near the gap region suggests weak hybridization at low energies, with Fe–Fe interactions governing the magnetic electronic structure. In contrast, the h-phase shows a significantly enhanced density of states at the Fermi level, indicating a more metallic character. The Fe $d$-states remain dominant but display: Broader features and increased overlap between majority (blue) and minority (red) components near $E_\text{F}$, with A reduced effective exchange splitting compared to the t-phase. This reflects a stronger itinerant character of the Fe $d$-electrons in the h-phase. Additionally, the S $p$-states exhibit slightly increased hybridization with Fe $d$-states across a wider energy range, contributing to the broader spectral features.

\newpage
\bibliographystyle{apsrev4-2}
\bibliography{Ref}